\documentclass[graybox]{svmult}

\usepackage{type1cm}        
\usepackage{makeidx}         
\usepackage{graphicx}        
\usepackage{multicol}        
\usepackage[bottom]{footmisc}

\usepackage{newtxtext}       %
\usepackage{newtxmath}       

\makeindex             

\usepackage{localdefs}  
\usepackage[numbers,sort&compress,longnamesfirst]{natbib} 

\usepackage{cleveref}
\DeclareMathOperator{\sinc}{sinc}

\begin{document}

\title*{Diffraction Gratings for X-ray Spectroscopy}
\author{Ralf K.~Heilmann, David P.~Huenemoerder, Jake A.~McCoy, and Randall L.~McEntaffer}
\institute{Ralf K.~Heilmann \at Space Nanotechnology Laboratory, MIT Kavli Institute for Astrophysics and Space Research, Massachusetts Institute of Technology, Cambridge, MA, USA, \email{ralf@space.mit.edu}
\and David P.~Huenemoerder \at MIT Kavli Institute for Astrophysics and Space Research, Massachusetts Institute of Technology, Cambridge, MA, USA, \email{dph@space.mit.edu}
\and Jake A.~McCoy \at Department of Astronomy and Astrophysics, the Pennsylvania State University, University Park, PA, USA, \email{jam1117@psu.edu}
\and Randall L.~McEntaffer \at Department of Astronomy and Astrophysics, the Pennsylvania State University, University Park, PA, USA, \email{rlm90@psu.edu}}
%
%
\maketitle


\abstract{
X-ray diffraction gratings play an essential role in high-resolution spectroscopy of astrophysical phenomena.  We present some scientific highlights from the \xray grating spectrometers (XGS) on board of the \chan and \xmm missions, XGS optical design, and the basic physics of grating diffraction geometry and efficiency.  We review design, fabrication, and performance of the currently orbiting transmission and reflection grating elements, followed by descriptions of the state-of-the art of more advanced grating technologies that promise orders-of-magnitude improvements in XGS performance, 
especially in combination with advanced \xray telescope mirrors.  A few key science questions that require new grating technology are posed, and powerful future mission concepts and recent and approved missions are presented.
}

\section{Introduction}
\label{sec:1}

Imaging of celestial objects, such as the planets of our solar system, stars, jets, nebulae and galaxies, 
inspires the popular
imagination. However, most imaging detectors are incapable of measuring the energy of the incident photons 
accurately and precisely enough
to draw clear-cut conclusions about the physical conditions and chemical abundances at the observed source.  
Much better results can be
achieved via dispersive (wavelength) or non-dispersive (energy) spectroscopy.  \xray detectors for the latter are
described in Chapter XX (\xray cryogenic detectors).  They usually feature a small, fixed energy resolution of a 
few eV, resulting in high energy resolution $E/\Delta E$ at higher \xray energies in the 6-10 keV range.  At lower 
energies (soft \xrays with energies $\sim 0.2 - 2$ keV), dispersive spectroscopy provides unsurpassed spectral 
resolving power ($R = \lambda/\Delta \lambda)$. In general, ``astronomy became astrophysics with the first 
spectrum. Spectroscopy determines compositions, magnetic field strength, space motion, rotation, multiplicity, 
planetary companions, surface structure, and other important physical traits'' \cite{2020Decadal}.  The soft \xray 
band, covering the characteristic lines of the most abundant ``metals'' (C, N, O, Ne, Mg, Si, S, Fe), is especially 
fertile ground for studying the composition and dynamics of the warm and hot, highly ionized plasmas that trace 
crucial feedback processes across up to fifteen orders of magnitude in spatial scale.  

In the visible part of the electromagnetic (EM) spectrum prisms are a well-know example of dispersive optics, splitting 
white light into a rainbow of colors in order of their wavelengths.  For \xrays the index of refraction for all 
matter is very close to one, and angular changes from refraction are very small.  In addition, all but the highest 
energy \xrays are easily absorbed upon transmission through millimeters or even micrometers of material.  
Dispersion of \xrays as a function of wavelength is therefore best achieved through diffraction from periodic 
structures.  This chapter focuses on periodic structures most useful in \xray astronomy, i.e., diffraction gratings with periodicity in one dimension (1D gratings).  

Diffraction gratings can be separated into two main categories: Transmission gratings (TG) and reflection gratings (RG).  
For transmission gratings photons are incident from one side of the grating, 
transmit though it, and disperse on the other side of the grating.  Reflection gratings disperse the incident photons back to the incident side.

\subsection{Scientific Highlights from the \chan and \xmm Grating Spectrometers}
\label{subsec:chandrascience}

Since the launches of \chan and \xmm in 1999 and their continuous
operation to the present, high dispersion soft \xray spectroscopy has
become routine through use of the High and Low Energy Transmission
Grating Spectrometers (HETGS \cite{HETG:2005} and LETGS \cite{Predehl_1992}) on the former, and the
Reflection Grating Spectrometer (RGS \cite{RGS}) on the latter.  ``High''
resolution refers both to the ability to {\em separate} neighboring
spectral features from one another, as well as the stronger sense of
{\em resolving} intrinsic feature profiles.  Both of these
capabilities have been fundamental for new fields of inquiry.  Here
we will highlight only few of the discoveries and areas of interest
illustrative of \chan and \xmm spectroscopy with the grating
instruments. A more detailed review of \xray astrophysics is given in
another chapter of this volume \citep[see][]{chap:fabian} (link to Fabian chapter on X-ray Astronomy).

In an early \chan/\hetg spectrum of the Classical T Tauri star, TW~Hya
\citep{Kastner:02}, the ability to distinguish individual emission lines of He-like species (\eli{O}{7}, \eli{Ne}{9}, \eli{Mg}{11}) revealed high-density plasma via the depression of forbidden lines.  This
discovery led to subsequent deeper observations of this and other
similar systems, and improved hydrodynamic shock modeling of
protostellar accretion \citep{brickhouse:al:2010, brickhouse:al:2012,
  argiroffi:maggio:al:2009, argiroffi:maggio:al:2012,
  argiroffi:drake:al:2017}.

Early \chan/\hetg spectra of hot stars also exploited the sensitivity
of the He-like line ratios, but this time to ultraviolet
photoexcitation instead of density, which allowed for localization of
the \xray emitting shocked plasma within the wind. Furthermore, with
wind velocities of order $1000\kms$ or more, the emission lines
themselves are resolved \citep{Waldron:Cassinelli:2001}.  The line
profiles, even though generated in only about $10\%$ of the wind
plasma, are key to diagnosing mass loss rates \citep{owocki:cohen:2001,
  Cohen:al:2006, oskinova:feldmeier:al:2006, naze:flores:al:2012},
  which is important for feedback of matter and energy on galactic scales.

Compact objects --- neutron stars (NS) and black holes (BH) --- in
\xray binaries (XRB) have been prime targets for high-resolution \xray
spectroscopy.  The high-energy emissions probe regions of extreme
conditions, which is crucial to understanding the physics of these
objects.  Often the central object or accretion disk emits a strong,
power law continuum, and against this, absorption lines form from the
intervening plasma.  The positions, widths, and variability reveal the
dynamics of matter in extreme conditions.  One such case is that of
GRO~J1655-40 \citep{miller:raymond:al:2008} in which there are many
absorption lines revealing a high-velocity outflow from an accretion
disk around a stellar-mass black hole, whose characteristics raise
questions about how the wind could be launched.  Her X-1 is a classic
case for study of accretion and outflow near a neutron star.  The
\xmm/\rgs spectra show a multitude of absorption and emission lines
which can be used to study the disk dynamics, geometry, and mass loss
rate \citep{kosec:fabian:al:2020}.

Compact objects are interesting in their own right, but they often
also provide a featureless continuum pencil-beam through the
interstellar medium (ISM).  This provides a mechanism for determining
the composition of the ISM, both elemental makeup and whether it is in 
a gaseous or solid state.  High resolution is needed to show the
complex absorption and emission components across the transition
edges
\citep{schulz:corrales:al:2016, lee:ogle:al:2001,rogantini:costantini:al:2020}.
Laboratory measurements are important to interpretation of edge
structure \citep{lee:2010,costantini:corrales:2022}.

Low-mass \xray binaries, cataclysmic variables, novae, 
and even some supernova remnants (if not too extended)
have also
been prime targets for high-resolution spectroscopy, especially at
longer wavelengths accessible to \xmm/\rgs or \chan/\letg, since these
sources often exhibit a soft state.  The recurrent nova, RS~Oph, is a
good example.  It has had two outbursts observed with \chan and \xmm,
which followed the evolution of emission and absorption lines,
studying both composition and dynamics \citep{ness:drake:al:2009,
  ness:beardmore:al:2023}.

At much larger scales, supermassive black holes are thought to reside
in the centers of Active Galactic Nuclei, where they give rise,
through poorly understood mechanisms, to relativistic outflows, greatly
impacting their environment to large distances. High-resolution \xray
spectra of these objects reveal these flows and their physical
conditions.  Variability is further used to constrain their geometry
and energetics. Example Spectra can be found in the review by \cite{laha:reynolds:al:2021}.

\begin{figure}[ht]
  \centering\leavevmode
  \includegraphics*[width=0.475\columnwidth]{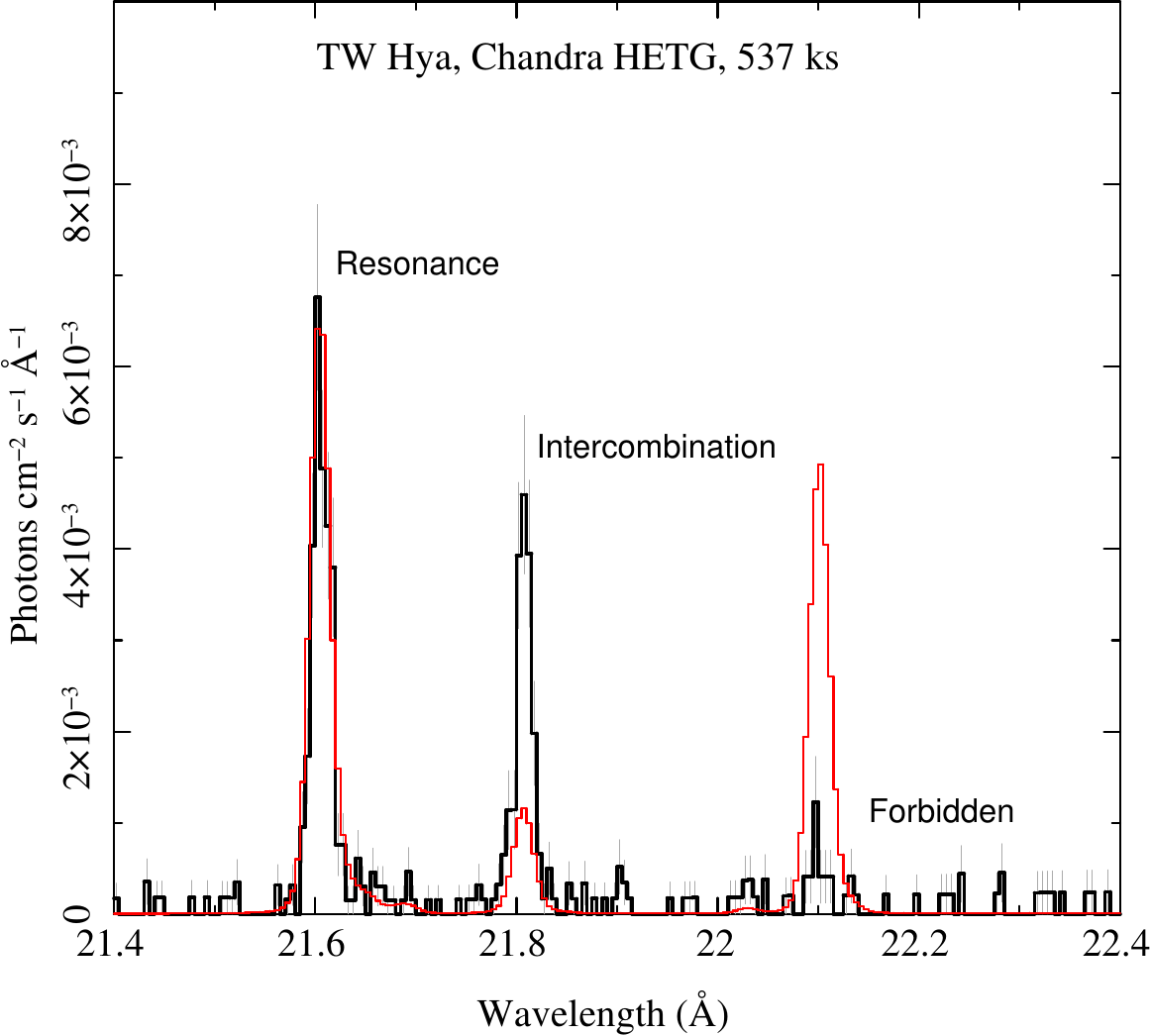}
  \includegraphics*[width=0.475\columnwidth]{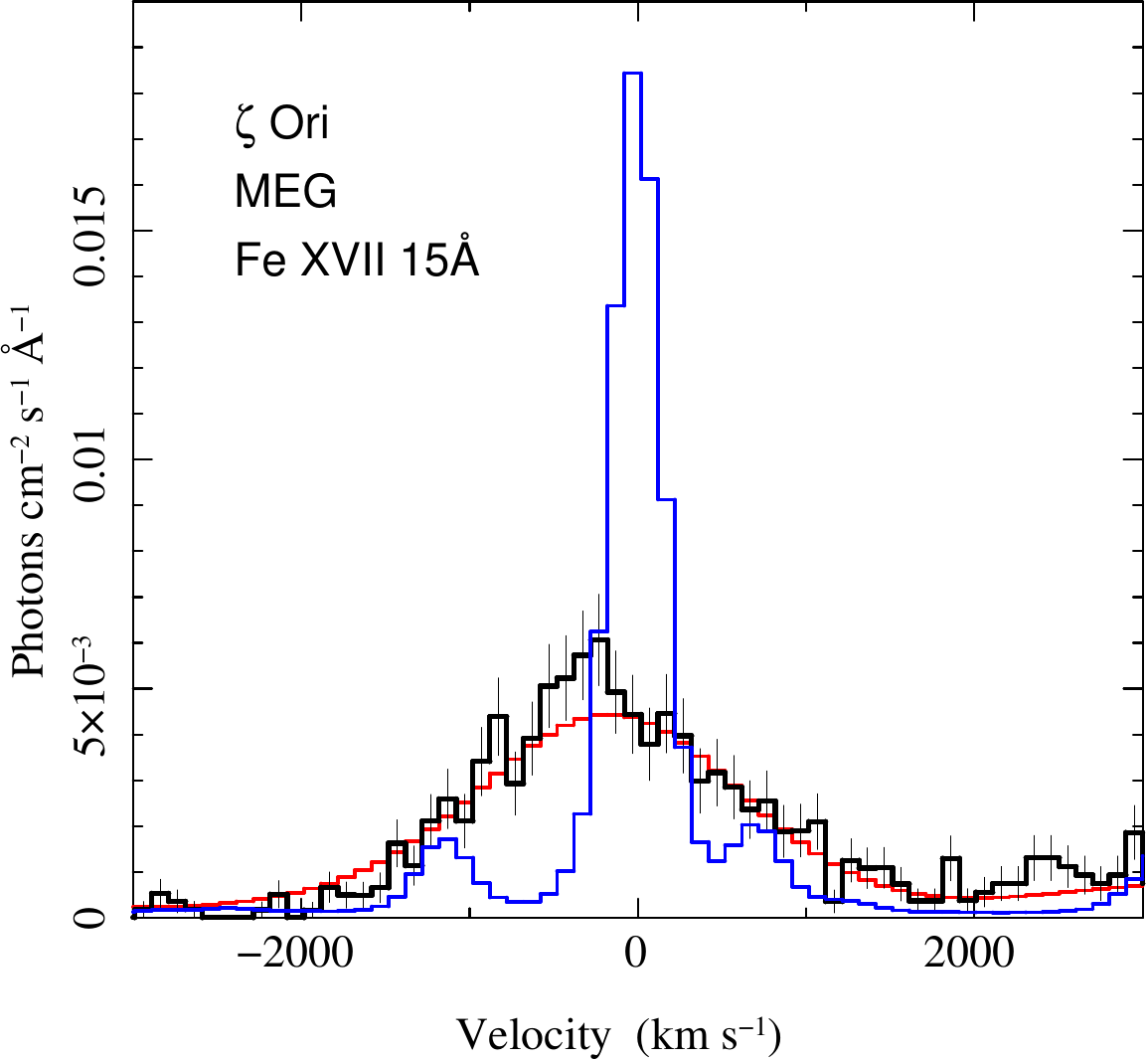}
  \includegraphics*[width=0.475\columnwidth]{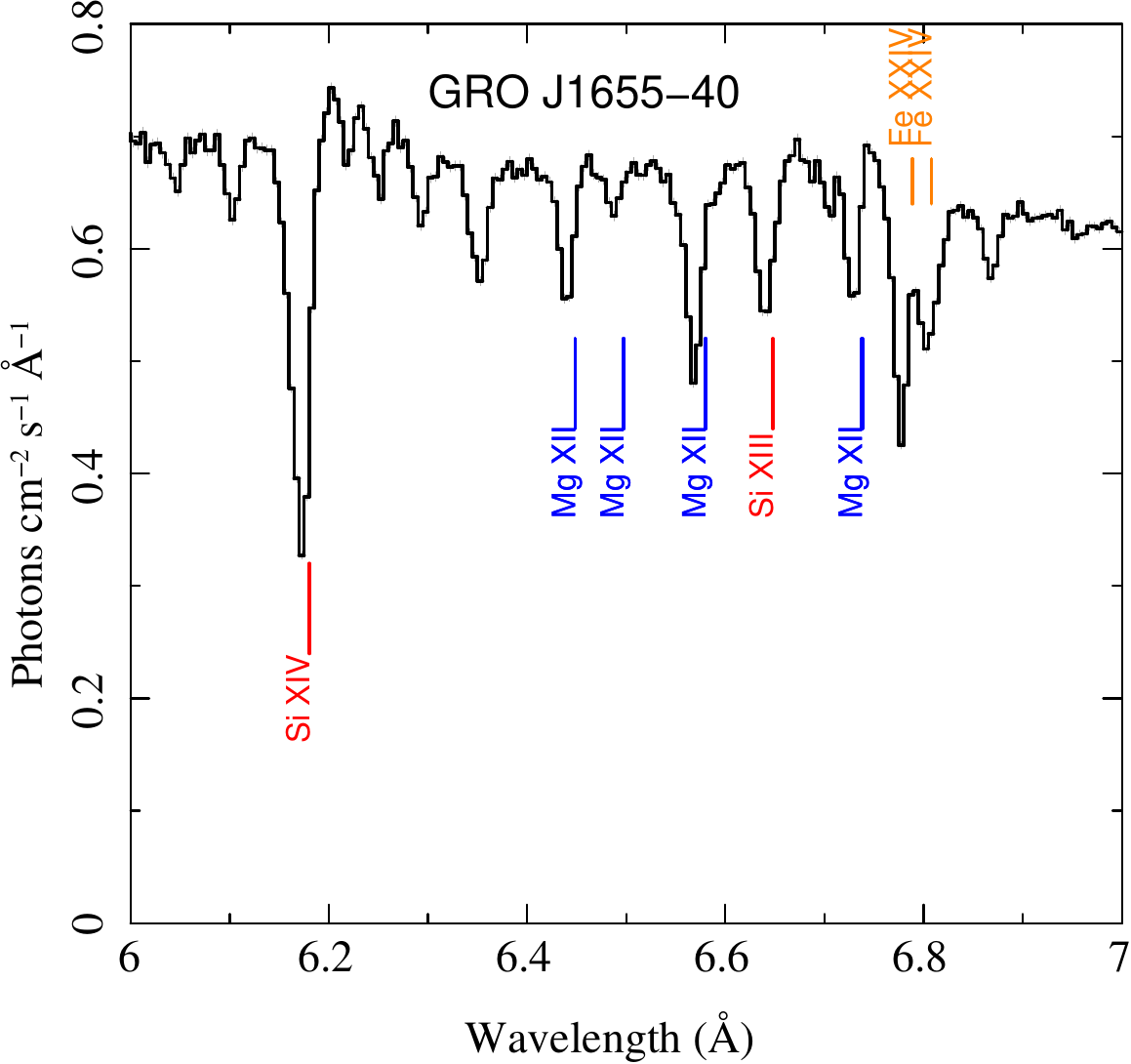}
   \includegraphics*[width=0.475\columnwidth]{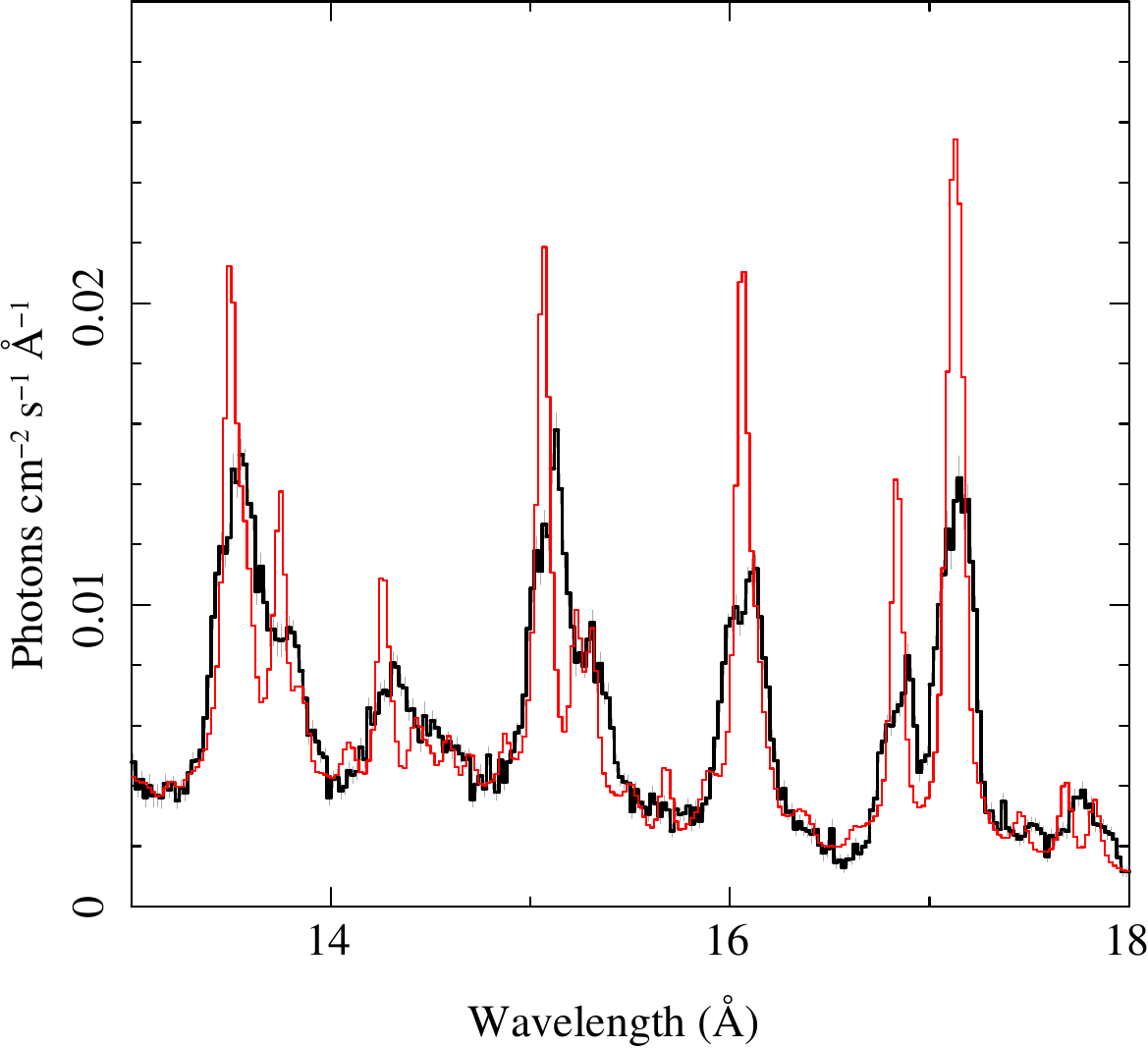}
  \caption{
    Upper left: Discovery of high-density plasma in TW Hya;
  black: observed HETG spectrum; red: low density model.  At high
  density, the forbidden line is weak and the intercombination line
  strong. 
  Upper right: black: the \eli{Fe}{17} $15\mang$ resolved emission line
  in the HETG spectrum of $\zeta\,$Ori. Red is a Gaussian-broadened plasma model  (which is
  not a good fit), blue shows the un-broadened model.  
  Lower left: the rich HETG absorption line spectrum of the BH XRB, GRO
  J1655-40, revealing broadened and shifted lines near a black hole.
  Lower right: A portion of an XMM/RGS spectrum of the supernova
  remnant, N132D, an extended source.  The black histogram shows the
  observed counts, and the red curve shows a model with the same flux, but
  at the instrumental resolution.  We can see additional broadening
  and structure in the observed spectrum which are due to the extended
  morphology of the source \citep[see][for more detail on this
  observation]{behar:al:2001}.
  }
  \label{fig:highlights}
\end{figure}

These are just a few examples of areas where high resolution \xray
spectroscopy has been crucial.  In Figure~\ref{fig:highlights} we show
sample spectra for some of these cases.


\section{Grating Spectrograph Optical Design}
\label{sec:GSOD}
Most \xray grating spectrographs (XGS) start with an \xray telescope that images objects at practically infinite distance onto a detector at the telescope focal plane.  
\xray telescopes need large apertures, ideally on the order of a square meter or more, to collect enough photons from faint sources in a reasonable amount of time. Individual \xray gratings of that size can not be fabricated with today's technology.
Instead, the telescope aperture is tiled with an aligned array of smaller gratings.

Typically, an objective grating array is placed in the converging beam of the telescope, such that diffracted \xrays of a given wavelength are focused into distinct diffraction orders.   
An imaging detector placed at the focal plane is used to record the dispersed spectra and to carry out separation of spatially overlapping orders from different wavelengths with its energy-dispersive response. 
The primary goal of the optical design is to maximize the spectral resolving power, $R$, over the wavelength band of interest. 
This is accomplished by placing the gratings as close as possible to the telescope mirrors and on the surface of a Rowland torus containing the telescope focus \cite{beuermann_aberrations_1978,Paerels10}.  
A sharp spectrum is then formed along the Rowland torus surface, assuming the individual grating dispersion directions are also aligned to each other.  

The resolving power for point sources usually is not limited by the optical design, but by the angular size of the telescope point-spread function (PSF) and the ability to blaze \xrays efficiently into high diffraction orders. 
With the grating array close to the telescope mirrors, $R$ 
can be estimated to first order as the ratio of diffraction angle over {the angular width of the} PSF. For telescopes with very small PSF such as \chan, $R$ will be smaller than this estimate due to small aberrations, fabrication limitations, and detector pixel sizes \cite{HETG:2005}.  Extended sources also can reduce $R$ 
since the effective
  resolving power is determined by the source image morphology
  convolved with the instrumental profile.  The relevant dimension is
  approximately the root-sum-of-squares of the instrumental full-width-half-max (FWHM) with
  the source FWHM \citep[for example, see Figure~3 in][]{dewey:2002}.
  If the source shape is known from imaging observations, then spatial
  profiles can be incorporated into the line response function.  Or, if
  there is a range of image orientations with respect to the dispersion
  direction (as in different spacecraft roll angles, or via positive and
  negative orders), then some spatial-spectral deconvolution can be
  done \citep{zhang:al:2019,suzuki:al:2020}.

The effective area of an \xray grating spectrograph is the product of the mirror effective area that feeds into the grating array, transmission through filters, the detector quantum efficiency, the fraction of the grating array not occupied by support structures or gaps, and the diffraction efficiency (DE) of the collected diffraction orders.

In \xray astronomy diffraction gratings are generally periodic along a single direction (1D gratings), described by a grating vector, which points from one grating groove to the next.  However, the grating vector can change as a function of position on the grating, depending on the choice of grating and diffraction geometry.

Diffraction of short wavelengths requires small grating periods to achieve useful diffraction angles.  For longer wavelengths astronomical diffraction gratings historically have been manufactured through mechanical ruling of aluminum blanks.  For \xray diffraction gratings the small features and high precision required are very difficult to achieve with mechanical means.  Consequently, the making of most modern \xray diffraction gratings involves lithographic patterning using light or electron beams to define the period progression of the grating.

In the following sections we describe various designs for transmission and reflection grating spectrographs in more detail.

\subsection{Diffraction Grating Framework}
\label{sec:diffgeom}

Here we present a brief, general introduction of grating diffraction in three dimensions, which is later applied to 
special geometries for transmission and reflection gratings.

\begin{figure}
 \centering
 \includegraphics[scale=0.5]{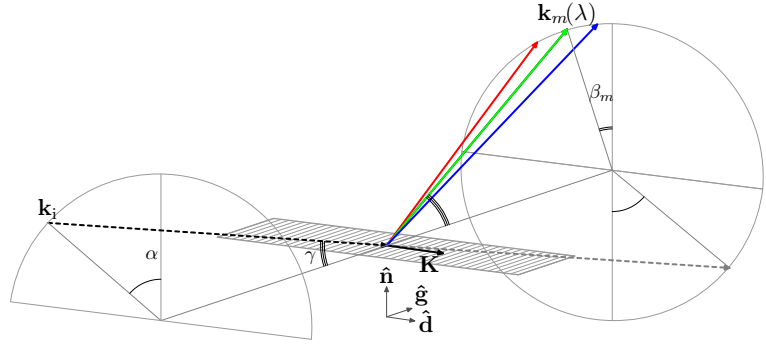}
 \caption{General scenario of a wavefront with wave vector $\mathbf{k}_{\text{i}}$ incident on a grating with a dispersion direction $\mathbf{\hat{d}}$, a surface-normal direction $\mathbf{\hat{n}}$ and a groove direction $\mathbf{\hat{g}}$. The $m^{\text{th}}$ propagating order with wave vector $\mathbf{k}_m$ can be reflective or transmissive while confined to a cone of opening angle $2 \gamma$. 
 Classical, in-plane diffraction is described by $\sin \left( \gamma \right) = 1$.}\label{fig:general_conical} 
 \end{figure}

We consider a plane containing a structure (substrate) with one-dimensional periodicity $p$ along the direction of unit vector $\mathbf{\hat{d}}$ (1D grating, see Fig.~\ref{fig:general_conical}).  The distance $p$ is the width of a grating groove.  The grating normal is $\mathbf{\hat{n}}$, the groove direction is $\mathbf{\hat{g}}$, and $\mathbf{\hat{d}} \times \mathbf{\hat{g}} = \mathbf{\hat{n}}$.

When a plane EM wave is incident on such a grating, some portion of the flux is reflected, some is transmitted, and some is absorbed.  Due to the periodic structure, the reflected and transmitted radiation is concentrated in a finite number of narrow angular ranges, the so-called diffraction orders.  There are many ways to derive the angles of these diffraction orders (path length difference, Huygens–Fresnel principle, etc.).  Here we follow a reciprocal-space-based approach that can also be applied to elastic electron and neutron scattering.  

The incident wave is represented by wave vector $\mathbf{k}_\text{i}$ with magnitude (wavenumber) $k \equiv |\mathbf{k}_\text{i}| = 2 \pi / \lambda$.

Next, we introduce the \emph{grating vector}, $\mathbf{K}$, which defines the spatial frequency of the grooves that make up the grating structure, as $\mathbf{K} \equiv 2 \pi \mathbf{\hat{d}}/p$. Its magnitude, $K$, is proportional to the groove density, $1/p$.  Depending on the choice of diffraction geometry, both the direction and magnitude of $\mathbf{K}$ can change with position on the grating for a variable-line-space or ``chirped" groove layout. In this section we limit discussion to the case where $\mathbf{K}$ is constant.

Due to the practically infinite mass of the grating compared to the small momentum of the incident wave, light is scattered elastically, i.e., $k$ is preserved.  The periodicity of the substrate allows for wave vector changes of integer multiples of the grating vector $\mathbf{K}$.  In compact form this can be written as \cite{RasmussenSPIE2004,FlanaganSPIE2004}

\begin{equation}\label{eq:compact_grating}
    \mathbf{\hat{n}} \times \left( \mathbf{k}_m - \mathbf{k}_{\text{i}} \right) = m K \mathbf{\hat{g}}  \quad \text{for } m = 0, \pm 1, \pm 2, \pm 3 \ldots ,
\end{equation}

\noindent
with $\mathbf{k}_\text{m}$ being the wave vector of the $m^{\mathrm {th}}$ diffraction order.

Cross products of \cref{eq:compact_grating} with $\mathbf{\hat{d}}$ and $\mathbf{\hat{g}}$ lead to two equations:
\begin{align}\label{eq:grating_dispersion_groove}
\left( \mathbf{k}_m - \mathbf{k}_{\text{i}} \right) \cdot \mathbf{\hat{d}} = m K  && {\mathrm {and}} && \left( \mathbf{k}_m -  \mathbf{k}_{\text{i}} \right) \cdot \mathbf{\hat{g}} = 0 .
\end{align}

This shows that the diffracted wave vectors $\mathbf{k}_m$ lie on a cone with $\mathbf{\hat{g}}$ as the axis of rotation with the same cone angle $\gamma$ as $\mathbf{k}_i$, and describes the general case of conical diffraction from a 1D grating.  In the grating normal direction we have 

\begin{equation}
    \left( \mathbf{k}_0 \pm \mathbf{k}_{\text{i}} \right) \cdot \mathbf{\hat{n}} = \pm 2 \mathbf{k}_{\text{i}} \cdot \mathbf{\hat{n}} ,
\end{equation}

\noindent
with the plus signs for the case of transmitted and the minus signs for the case of reflected diffraction orders.  The propagating orders must obey 
$|\mathbf{k}_m \cdot \mathbf{\hat{d}}| < |\mathbf{k}_i \times \mathbf{\hat{g}}|$, which limits the range of $m$. 

The wave-vector relations defined by \cref{eq:grating_dispersion_groove} can be written in a more familiar form by setting $\mathbf{\hat{d}} = \mathbf{\hat{x}}$, $\mathbf{\hat{n}} = \mathbf{\hat{y}}$ and $\mathbf{\hat{g}} = - \mathbf{\hat{z}}$.
Inserting wave vectors 
\begin{align}
\mathbf{k}_\text{i} = k_{x,\text{i}} \mathbf{\hat{x}} + k_{y,\text{i}} \mathbf{\hat{y}} + k_{z,\text{i}} \mathbf{\hat{z}} &&
{\mathrm {and}} &&
\mathbf{k}_m = k_{x,m} \mathbf{\hat{x}} + k_{y,m} \mathbf{\hat{y}} + k_{z,m} \mathbf{\hat{z}} 
\end{align}
into \cref{eq:grating_dispersion_groove} yields 
\begin{align}
k_{x,m} = k_{x,\text{i}} + m K && {\mathrm {and}} && k_{z,m} = k_{z,\text{i}}. 
\end{align}

Respectively, these equations describe diffraction in the $x$-direction, with diffracted orders arising from integer multiples of $K$ added to $k_{x,\text{i}}$, and specular reflection or direct transmission in the $z$-direction, with $k_{z,\text{i}} \equiv k \cos \left( \gamma \right)$ being preserved for all $m$. 
The vectors $\mathbf{k}_m$ and $\mathbf{k}_\text{i}$ with $k_{z,m} = k_{z,\text{i}}$ can be written in terms of spherical coordinates as 
\begin{align}
    \begin{split}
        \mathbf{k}_\text{i} &= k \left[ -\sin \left( \alpha \right) \sin \left( \gamma \right) \mathbf{\hat{x}} -\cos \left( \alpha \right) \sin \left( \gamma \right) \mathbf{\hat{y}} + \cos \left( \gamma \right) \mathbf{\hat{z}}  \right] \\
        \mathbf{k}_m &= k \left[ \sin \left( \beta_m \right) \sin \left( \gamma \right) \mathbf{\hat{x}} + \cos \left( \beta_m \right) \sin \left( \gamma \right) \mathbf{\hat{y}} + \cos \left( \gamma \right) \mathbf{\hat{z}} \right] ,
    \end{split}
\end{align}
where $\alpha$ is the azimuthal incidence angle and $\gamma$ is the angle between $\mathbf{k}_\text{i}$ and $\mathbf{\hat{g}}$. 
The relation $k_{x,m} = k_{x,\text{i}} + m K$ then yields the \emph{generalized grating equation} that describes conical diffraction: 
\begin{equation}\label{eq:general_GE}
    \sin \left( \alpha \right) + \sin \left( \beta_m \right) = \frac{m \lambda}{p \sin \left( \gamma \right)} \quad \text{for } m = 0, \pm 1, \pm 2, \pm 3 \ldots ,
\end{equation}
where the special case of in-plane diffraction has $\sin \left( \gamma \right) = 1$ with $k_{z,\text{i}} = 0$ such that the cone of diffraction opens fully into a 2D arc of diffraction in the $xy$-plane. 
In a combined RG and TG framework, 
$\beta_m$ spans $2 \pi$ with $\cos \left( \beta_m \right) > 0$ and $\cos \left( \beta_m \right) < 0$ describing reflected and transmitted orders, respectively. 
However, it is sometimes convenient to redefine $\beta_m \to \beta_m - \pi$ for transmitted orders such that the sign of $\sin \left( \beta_m \right)$ in \cref{eq:general_GE} flips. 
TGs have to be thin enough to have useful transmission, and refraction can be neglected.

The grating equation (\ref{eq:general_GE}) does not describe how the intensity of the incident \xray flux is distributed over the various diffraction orders $m$.  It is the electron density distribution of the grating groove that leads to phase shifts and absorption and controls the relative intensities of diffraction orders as a function of $\lambda$, $\alpha$, and $\gamma$.  

For \xrays we can assume that the grating structure dimensions are much larger than $\lambda$.  In the following, we also neglect polarization effects for simplicity, which is often justified at near-normal or grazing angles of incidence \cite{Attwood17}.  Long \xray telescope focal lengths place detectors in the far field.  Under these combined conditions scalar Kirchhoff diffraction theory can be applied in the Fraunhofer limit.  The DE in the $m^{\mathrm {th}}$ order can then be written as \cite{BornWolf,Michette,HETG:2005}

\begin{equation} \label{eq:general_DE}
    {\frac{I^{(m)}} {I_0}} = |F(\mathbf{k}_m - \mathbf{k}_i)|^2 \left( \frac{ \sin (Ns)} {N \sin (s)}  \right) ^2,
\end{equation}

\noindent
where $I_0$ is the flux incident on the grating,
$s \equiv \left[\sin \left( \alpha \right) + \sin \left( \beta_m \right)\right] \sin \left( \gamma \right) p \pi /\lambda$,
and $N$ is the number of illuminated grating grooves.  $F(\mathbf{k}_m - \mathbf{k}_i)$ is the Fourier transform of the groove function $f(x)$ that describes the path length and phase shift that \xrays undergo as they pass through or get reflected from a grating groove:

\begin{equation}
    F(\mathbf{k}_m - \mathbf{k}_i) = {\left(1/p \right)} \int_0^p f(x) 
     \exp{\left( -i \left( \mathbf{k}_m - \mathbf{k}_{\text{i}} \right) \cdot \mathbf{\hat{d}} x \right) } dx .
\end{equation}

The groove function can be written as 

\begin{equation}\label{eq:groove_func}
    f(x) = \exp{\left\{ ik\left[ n(k) - 1 \right] y\left( x/p\right) \right\}} ,
\end{equation}

\noindent
where $n(k) = 1 - \delta_n (k) + i\beta_n (k)$ is the complex index of refraction of the grating material ($\delta_n (k), \beta_n (k) \ll 1$) \cite{Attwood17}, and $y(x/p)$ describes the path length through the grating material (TG) or upon reflection (RG). 

The second factor in Eq.~\ref{eq:general_DE} peaks near angles given by the grating equation (\ref{eq:general_GE}) and converges into a comb of Dirac delta functions at $s = m\pi$ for $N \rightarrow \infty$.  

\begin{figure}
 \centering
 \includegraphics[scale=0.43]{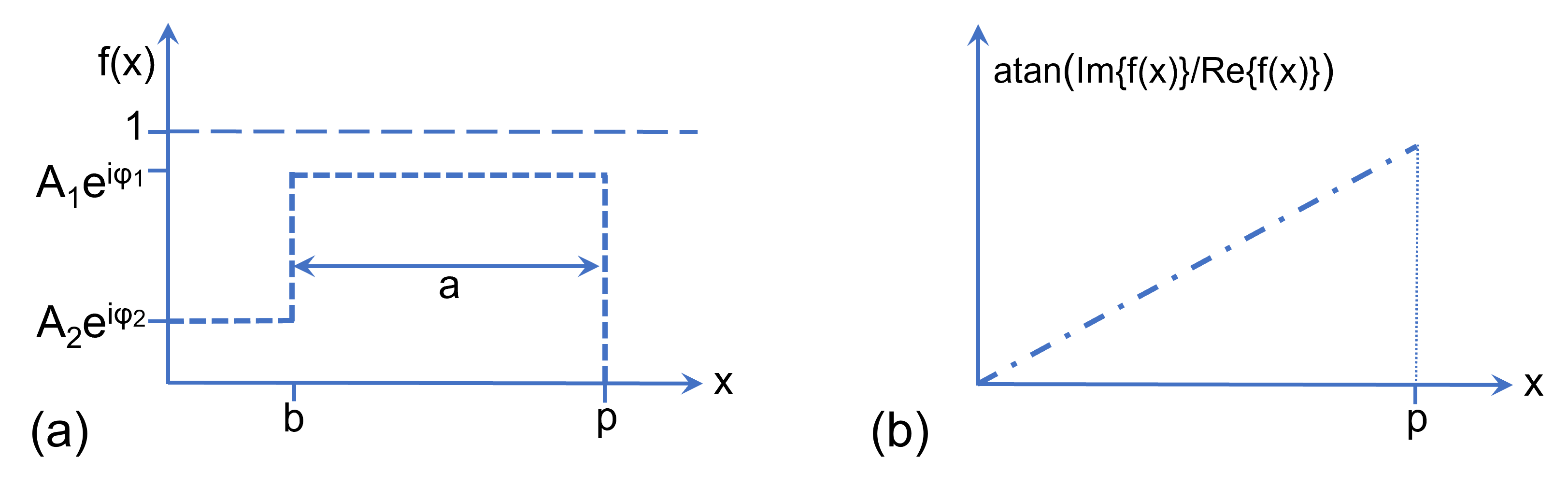}
 \caption{(a) Groove function $f(x)$ for a lamellar grating profile.  (b) Phase progression of $f(x)$ for an ideal blazed grating.  See following sections for details.}\label{fig:groovefunction} 
 \end{figure} 

Some specific cases of $f(x)$ are discussed in the following sections.

\section{Transmission Gratings}
\label{sec:TG}

For transmission gratings \xrays are incident on one side of the grating and exit on the other side.  
For all gratings considered in this section, \xrays are incident close to the grating normal and lie (to first order) in the plane defined by the grating normal and the grating vector ($\gamma \approx \pi /2$).  
The grating equation simplifies to its well-known in-plane version \cite{BornWolf}

\begin{equation}
{\frac{m \lambda}{p}} = \sin (\alpha) - \sin (\beta_m) .
\label{eq:ipge}
\end{equation}

\noindent
To achieve useful diffraction angles 
(assuming $R > 1000 \approx |m| \lambda /(\mathrm {PSF} p)$, and PSF $\sim 10^{-5} \mathrm { \ rad} \sim$ arcsec level), the grating period $p$ should be $\lessapprox 10^2$ $|m|\lambda$, which is a manufacturing challenge for \xrays.

The advantages of transmission gratings are that they are relatively insensitive to misalignments due to a lack of reflection, very thin (low mass), and operate near normal incidence (which relaxes their size requirements compared to grazing-incidence RGs).

Many TGs can be approximated to have a rectangular groove function, where $f(x) = A_1(k) \exp{\left(i\phi_1(k)\right)}$ for $b < x < p$ and $f(x) = A_2(k) \exp{\left(i\phi_2(k)\right)}$ for $0 < x < b$, and where $A_i$ describes the transmission and $\phi_i$ the phase shift for region $i$ (see Fig.~\ref{fig:groovefunction}(a)).
One then obtains for the DE of the $m = 0$ order

\begin{equation}
{\frac{I^{(0)}} {I_0}} = \left[ (p-b)^2 A_1(k)^2 + b^2 A_2(k)^2 + 2 b (p - b) A_1(k)A_2(k)\cos \left( \phi_2(k) - \phi_1(k) \right) \right] ,
\label{eq:0th}
\end{equation}

and for the $m^{\mathrm {th}}$ order

\begin{equation}
{\frac{I^{(m)}} {I_0}} = \left[ A_1(k)^2 + A_2(k)^2 - 2 A_1(k)A_2(k)\cos \left( \phi_2(k) - \phi_1(k) \right) \right]
\left( \frac{ \sin (m \pi b/p)} {m \pi}  \right) ^2.
\label{eq:mth}
\end{equation}

The DE can be minimized for $0^{\mathrm {th}}$ order and maximized for non-zero orders if $b = p/2$, and if $\phi_2(k) - \phi_1(k) = \pi$ can be achieved.

For non-normal incidence the groove function for rectangular bars takes on a trapezoidal shape, similar to the bar thickness projected onto the incident wavefront.

For the simple case of an ideal zero-thickness grating with gap width $a = p-b$ between fully opaque grating lines, $f(x) = 1$ for $b < x < p$ and zero elsewhere. This is a good approximation of the case of grating bars with thickness $d$ greater than the absorption length for a given wavelength.

The DEs are then independent of wavelength, with the $0^{\mathrm {th}}$ order DE equal to $(1 - b/d)^2$, and the $m^{\mathrm {th}}$ order DE given by

\begin{equation}
{\frac{I^{(m)}} {I_0}} = \left( \frac{ \sin (m \pi b/p)} {m \pi}  \right) ^2,
\label{eq:opaque_mth}
\end{equation}

\noindent
where the efficiency in $\pm 1{\mathrm {st}}$ orders peaks at 10.13\% for $b=0.5p$, and even orders are extinct. 
The challenge in \xray TG design and fabrication lies in the small grating periods, the weak phase shifts, the easy absorption of \xrays, and the difficulty of diffracting efficiently at large angles.

In the following subsections we first describe two complementary transmission grating spectrographs aboard NASA's \chan \ \xray observatory, which launched in 1999 and still is in operation \cite{HETG:2005}. We then discuss a novel and more powerful TG design that has been under development for over ten years and has been proposed for future \xray missions.

\subsection{The \chan Low Energy Transmission Grating Spectrometer (LETGS)}
\label{subsec:LEG}

The \chan \xray Observatory \cite{Chandrabook}, formerly known as {\it AXAF}, consists of focusing optics, two retractable grating arrays, and two interchangeable readout cameras.  The focusing optics are four nested, co-aligned, full-shell, 10-m focal length Wolter-I grazing incidence mirrors made from Zerodur and coated with iridium.  The combined mirror PSF, including camera effects, is about 0.5 arcsec FWHM.  The shell diameters range from about 0.63 to 1.2 m.

The LETGS was designed for high-resolution spectroscopy for soft \xrays up to 175 \AA \ in wavelength.  The LETGS objective grating array consists of 540 round grating facets, mounted to a roughly $1.1$ m-diameter Al frame that can be inserted into the telescope beam and places the facets on the surface of a Rowland torus about 8.6 m from the telescope focus.  The facets are arranged in four concentric circles, each covering the \xrays exiting from one of the mirror shells.  Each grating facet consists of a freestanding Au diffraction-grating structure with $p = 991$ nm, $a = p-b \sim 457$ nm, and grating bar thickness (along the transmission direction) $d = 474$ nm.  The bars are held in place by a fine Au cross-support structure with $p = 25.4$ $\mu$m and $d = 2.5$ $\mu$m.  The fine-support structure is in turn tethered to a triangular coarse-support Au structure of 2 mm pitch.  The combined support structures obscure less than 20 \% of the grating area \cite{Chandrapropguide}.  

\begin{figure}[ht]
  \centering
  \includegraphics*[width=0.36\columnwidth]{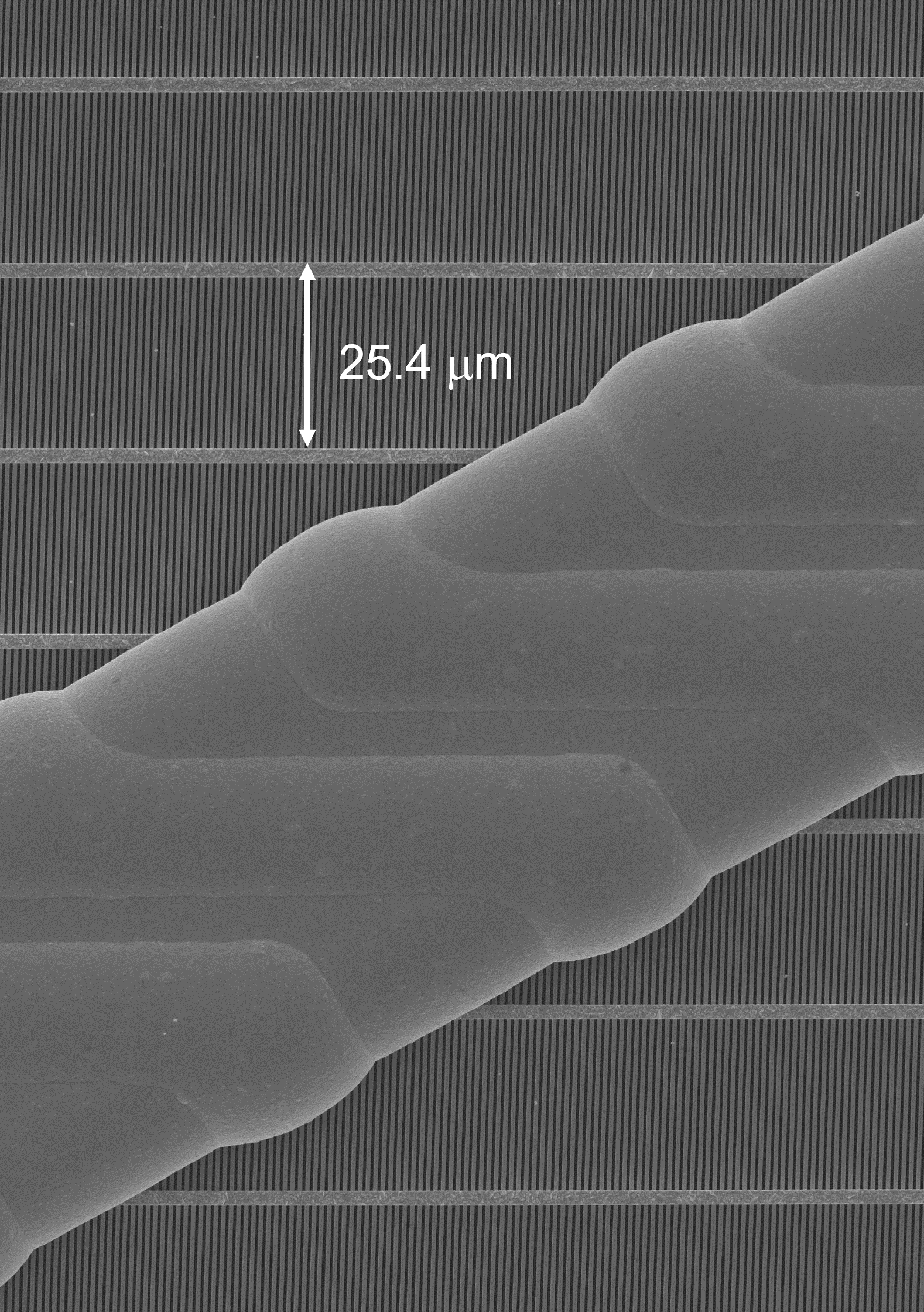}
  \includegraphics*[width=0.34\columnwidth]{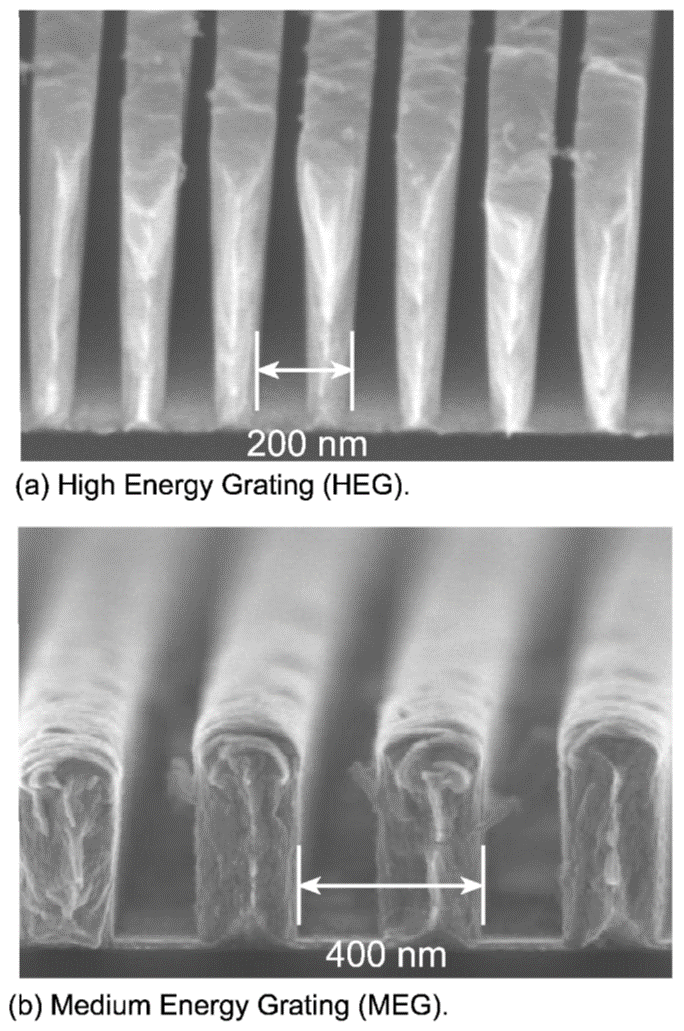}
  \includegraphics*[width=0.175\columnwidth]{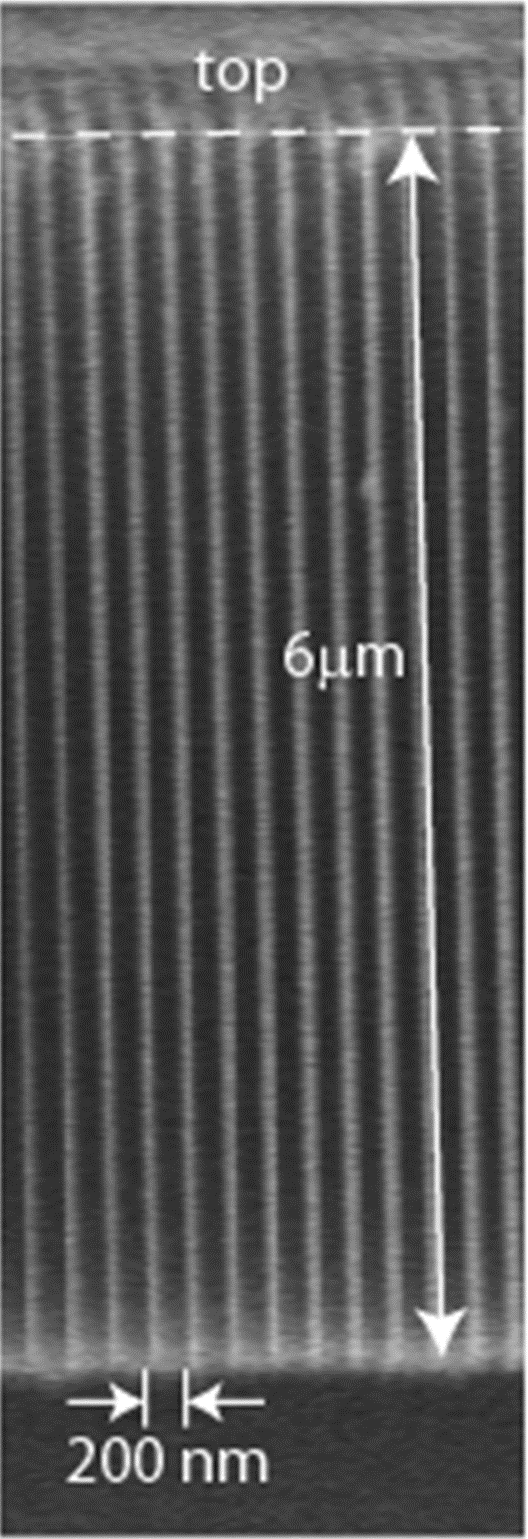}
 \caption{SEM images of TGs.  Left: Top-down view of LETG, showing large triangular mesh bar, cross supports, and fine grating bars, all freestanding and made of Au.  Middle top: Cleaved cross section of Au HEG on polyimide membrane.  Middle bottom: Cleaved cross section of Au MEG on polyimide membrane.  Right: Cleaved cross section of Si CAT grating after front side DRIE and KOH polish.  A cross support bar is visible in the background on top.  See text for more details.
  }
 \label{fig:TG_SEMs}
 \end{figure}

\begin{figure}[ht]
  \centering
  \includegraphics*[width=0.98\columnwidth]{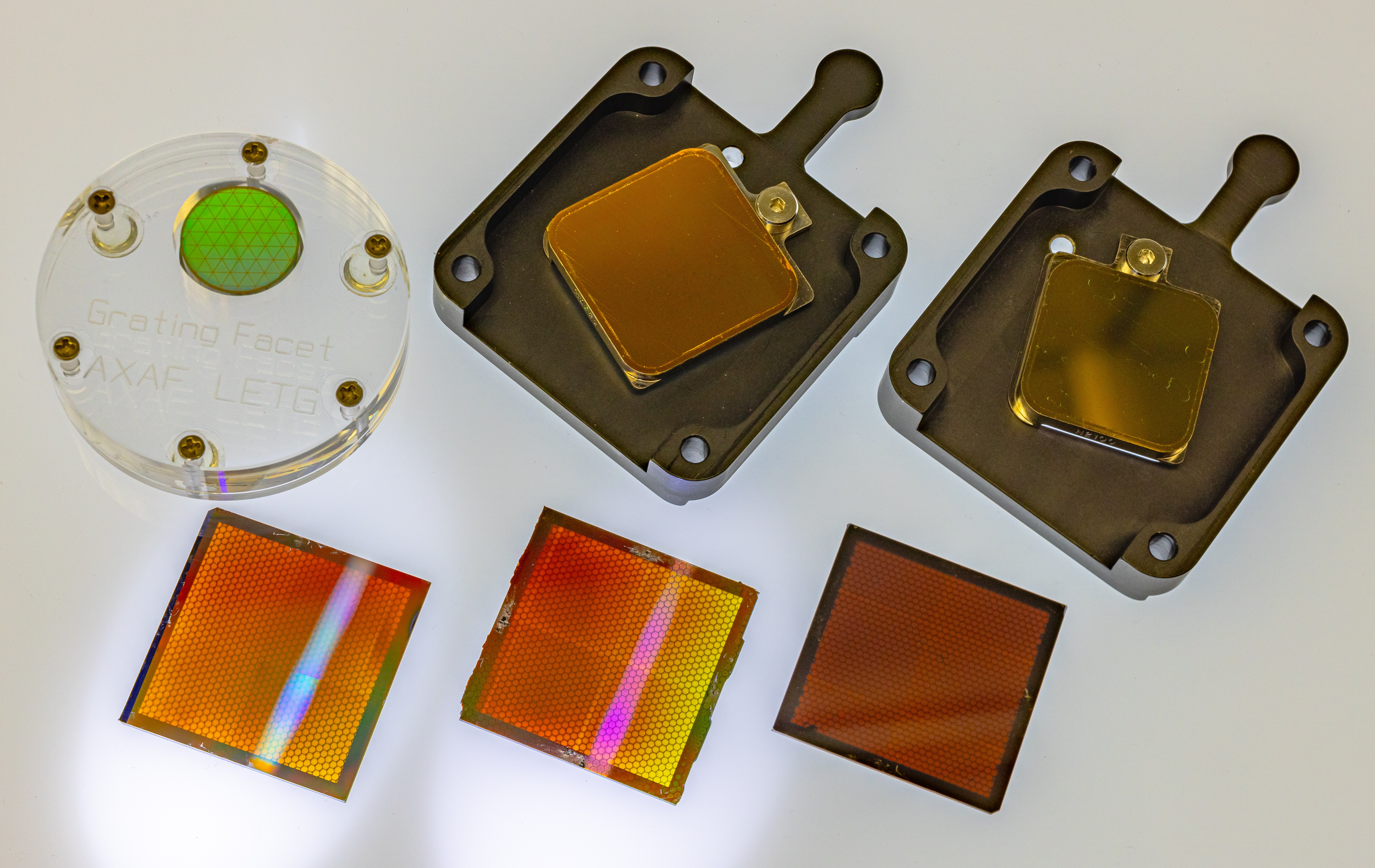}
  \caption{TGs, clockwise from top left: LETG (in display case), MEG and HEG (each mounted to a handle for \xray testing), three CAT gratings with \arcus dimensions.  Visible diffraction stems from the 5 $\mu$m-period support bars. See text for more details. (Image credit: G. Furesz.)}
  \label{fig:TGfam}
\end{figure}

All three different structures were started with mechanical diamond-tip ruling of a metal film-coated glass substrate; one for each structure.  From each original master, multiple working masters were produced via photolithograpy.  The final gratings start from a substrate covered with a thin conducting layer and a photoresist layer.  The working master is used to expose the photoresist, and after development the removed areas are plated up in an electrolyte bath, forming the grating bars.  The process is repeated with each of the cross-support structure masters and additional photoresist layers \cite{Predehl_1992}.  After dissolution of the conducting base layer the grating structure can be removed from the substrate and glued to a rigid, 16 mm-diameter circular frame.  Fig.~\ref{fig:TG_SEMs} shows a scanning electron microscope (SEM) image with all three levels of the freestanding Au structure, and Fig.~\ref{fig:TGfam} shows a round TG facet.  The gratings are mounted to Al modules in sets of three, and the modules are mounted to the insertable Al frame \cite{Braun_AO_1979,Predehl_1992}.

\begin{figure}
 \centering
 \includegraphics[scale=0.45]{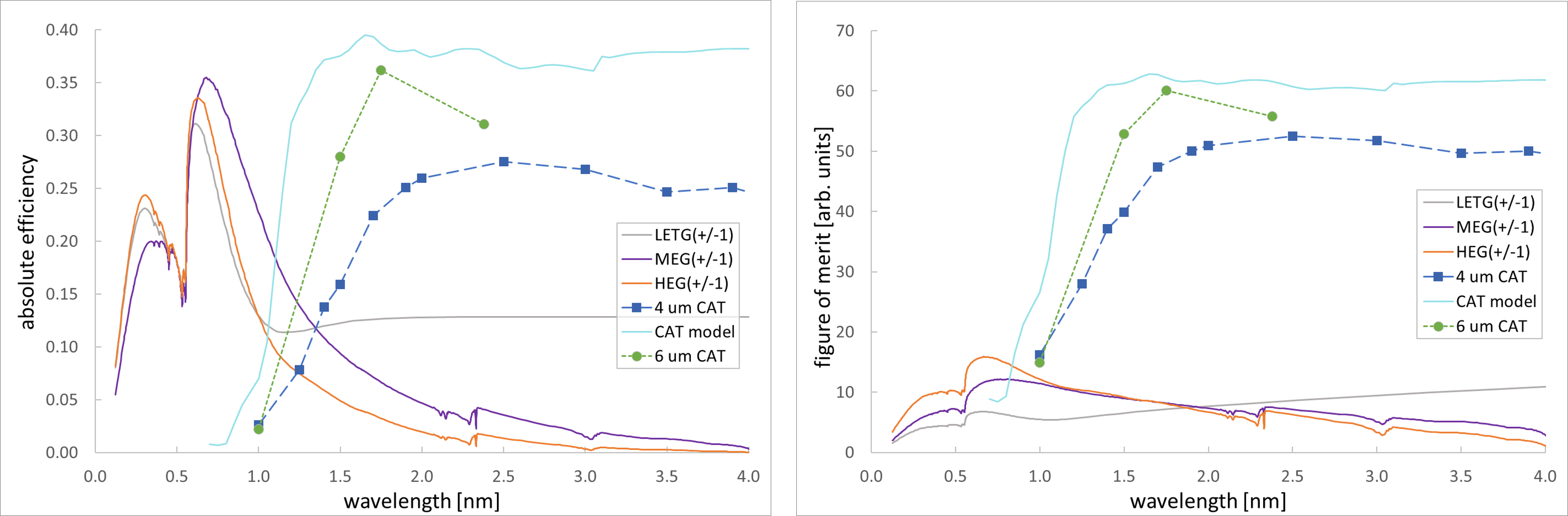}
 \caption{
 Comparison of TG diffraction efficiencies and figure of merit.  {\bf Left:} For \chan gratings the sum of $\pm 1^{\mathrm {st}}$ order DEs from calibration files is shown \cite{marshall:2012}.  These DEs include all losses from support structures and gaps between facets.  For CAT gratings all orders within the blaze envelope are summed up, and a total loss of 31\% from support structures and gaps is assumed. ``4 $\mu$m CAT" and ``6 $\mu$m CAT" are from synchrotron measurements of CAT gratings with $d = 4$ and 6 $\mu$m.  The latter are from the first prototypes of this thickness.  ``CAT model" is the sum of blazed order DEs from a model with $d = 6$ $\mu$m, $b = p - a = 40$ nm, and 1.5 nm sidewall roughness.
 {\bf Right:} Figure of merit for weak line detection 
 ($\sim \sqrt{A_\mathrm{eff} R}$).  The same mirror effective area and constant detector efficiency is assumed for all plots.  For a CATXGS, $R = 10^4$ is assumed, based on conservative lab measurements \cite{SPIE2016,AO2019,ApJ}. The HETG and LETG instruments complement each other over the shown range.  A CATXGS is superior by far for $\lambda > 1$ nm.  (See text for more details.) }
 \label{fig:TG_DE}
 \end{figure}

The LETGS delivers $R > 1000$ for $\lambda > 50$ \AA \ \cite{Chandrapropguide}.    For shorter wavelengths the thickness $d$ and width $b$ of the Au bars are optimized to take advantage of variations in the Au index of refraction and absorption edges, using partial transparency of the Au bars and near-$\pi$ phase shifts upon transmission to minimize $0^{\mathrm {th}}$ and other even orders and maximize $\pm 1^{\mathrm {st}}$ order efficiencies around 20\% per order near 6 \AA \ \cite{henkeTG} (see Eq.~\ref{eq:mth}).  Above $\sim 20$ \AA \ transmission through the Au bars is negligible, and the LETG DE closely follows Eq.~\ref{eq:opaque_mth} (see Fig~\ref{fig:TG_DE}).

For the comparison of different spectroscopic instruments it is useful to look at figures of merit (FOM) for different measurement goals.  An example is the FOM for weak line detection, given as $\sqrt{A_\mathrm{eff} R}$ (also shown in Fig~\ref{fig:TG_DE}), where $A_\mathrm{eff}$ is the effective area of the instrument \cite{kaastra:2017}.


\subsection{The \chan High and Medium Energy Transmission Gratings}
\label{subsec:HEGMEG}

The \chan High-Energy Transmission Grating Spectrometer (HETGS) employs two different types of gratings: 144 high-energy (HEG) and 192 medium energy (MEG) gratings \cite{HETG:2005}.  They are optimized for a combination of higher dispersion in first order (200 nm and 400 nm grating period, respectively) and higher DE in the 2-8 keV energy range. The gratings are arranged similar to the LETG gratings, but on a separate retractable Al frame.  When inserted into the beam, the MEGs cover the outer two mirror shells, while the HEGs cover the inner two shells.  The two sets of gratings are oriented with their dispersion axes offset $10^{\circ}$ from each other, producing spatially separated spectra on the detector.  Most alignment degrees of freedom were controlled using machining tolerances.  The dispersion axes (rotation around $\mathbf{\hat{n}}$) were aligned using the polarizing grating properties for visible light \cite{Anderson}.

The challenge in fabricating these gratings was to combine the phase shifting properties of dense metals to increase DE similar to the LETG, but with smaller grating periods that diffract \xrays to larger $|\beta_m|$.  Higher line density, $1/p$, makes mechanical ruling infeasible, and $\pi$ phase shifts require grating bars with higher aspect ratios.

The small periods are achieved using UV laser interference lithography at 351 nm wavelength \cite{IL-SPIE1994} in the initial patterning of a photoresist layer on top of a complex layer stack grown on Si wafers.  
The developed grating pattern
is transferred into a polymer layer that has a thickness close to the final height of the grating bars.  To create the grating bars, Au is plated between the polymer bars, and the polymer is then removed, leaving 120 nm (HEG) and 208 nm wide (MEG) Au bars with average thickness of 510 nm (HEG) and 360 nm (MEG), on a 25 nm thin Au/Cr plating base on top of a 980 nm (HEG) or 550 nm thick (MEG) polyimid layer (see Fig.~\ref{fig:TG_SEMs}).  Finally, a portion of the supporting wafer is etched away, and the now freestanding polyimid membrane is bonded to an invar flight frame \cite{nanogiga} (see Fig.~\ref{fig:TGfam}).

The resulting grating facets feature fine-period Au grating bars without absorbing coarse support structures, but the supporting polyimid membrane absorbs some soft \xrays, leading to lower DE at longer $\lambda$ (see Fig.~\ref{fig:TG_DE}).  At shorter $\lambda$, the HETGs also act as phase-shifting TGs, peaking with almost 20\% $\pm 1^{\mathrm {st}}$ DE near 6.5-7 \AA \ , and then again up to 14\% per order near 3-4 \AA \ .  The detailed shape of the efficiency curve can be modeled well using a multivertex model for the grating bar cross section instead of a simple rectangle \cite{HETG:2005}.  The resolving power for the HEG spectrum exceeds 1000 for $\lambda > 9$ \AA \ and $\lambda > 17$ \AA \ for the MEG spectrum in $\pm 1^{\mathrm {st}}$ orders.

\subsection{Critical-Angle Transmission Gratings}
\label{subsec:CAT}

The \chan HETGs are tuned for good DE below 1 nm (above $\sim 1.25$ keV), but become very inefficient at lower energies due to the absorbing polyimid membranes.  The HETGS resolving power is highest at low energies and decreases at higher energies since DE peaks in $\pm 1^{\mathrm {st}}$ orders.  The LETG has better DE at long $\lambda$, but lower $R$.  A better TG would have high DE in high $|m|$ (larger $|\beta_m|$, larger $R$) across a broad bandpass.  

Microcalorimeter technology has made great strides in the last decades and can provide high energy resolution $E/\Delta E$.  The Resolve instrument on the XRISM mission \cite{Resolve2022} has $\Delta E$ near 5 eV, and ground-based technology has shown $\Delta E$ below 1 eV (see Chapter on \xray cryogenic detectors...).  However, at energies below 1 keV the energy resolution remains modest compared to improved grating spectrometer designs, which emphasize performance for $\lambda \gtrapprox 1$~nm. 

\begin{figure}[ht]
  \centering
  \includegraphics*[width=0.4\columnwidth]{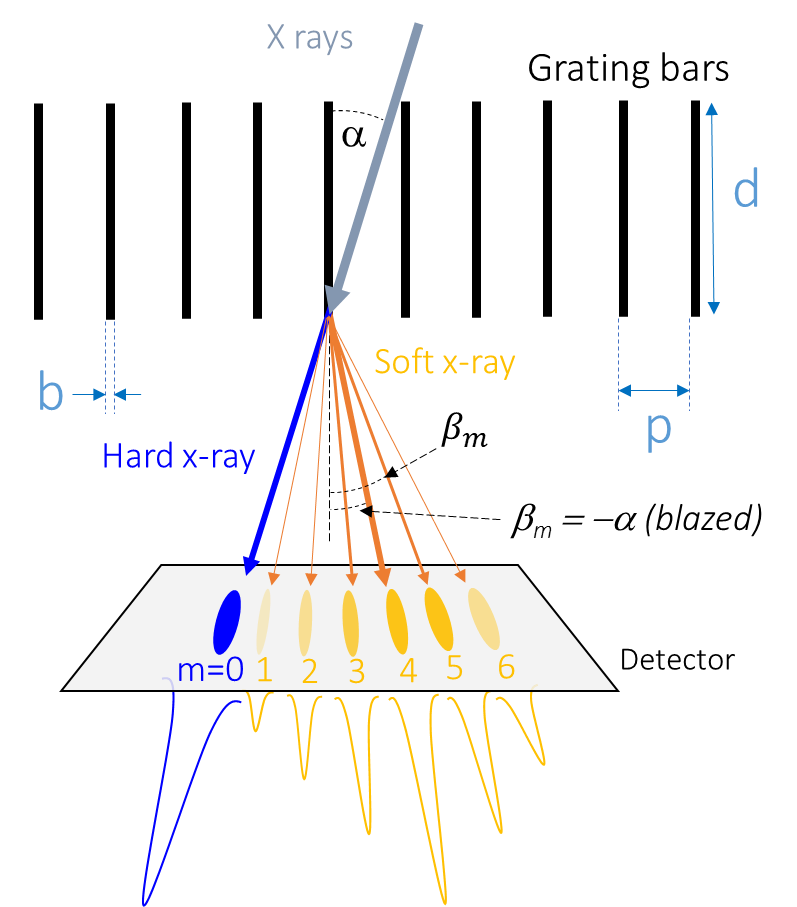}
  \includegraphics*[width=0.55\columnwidth]{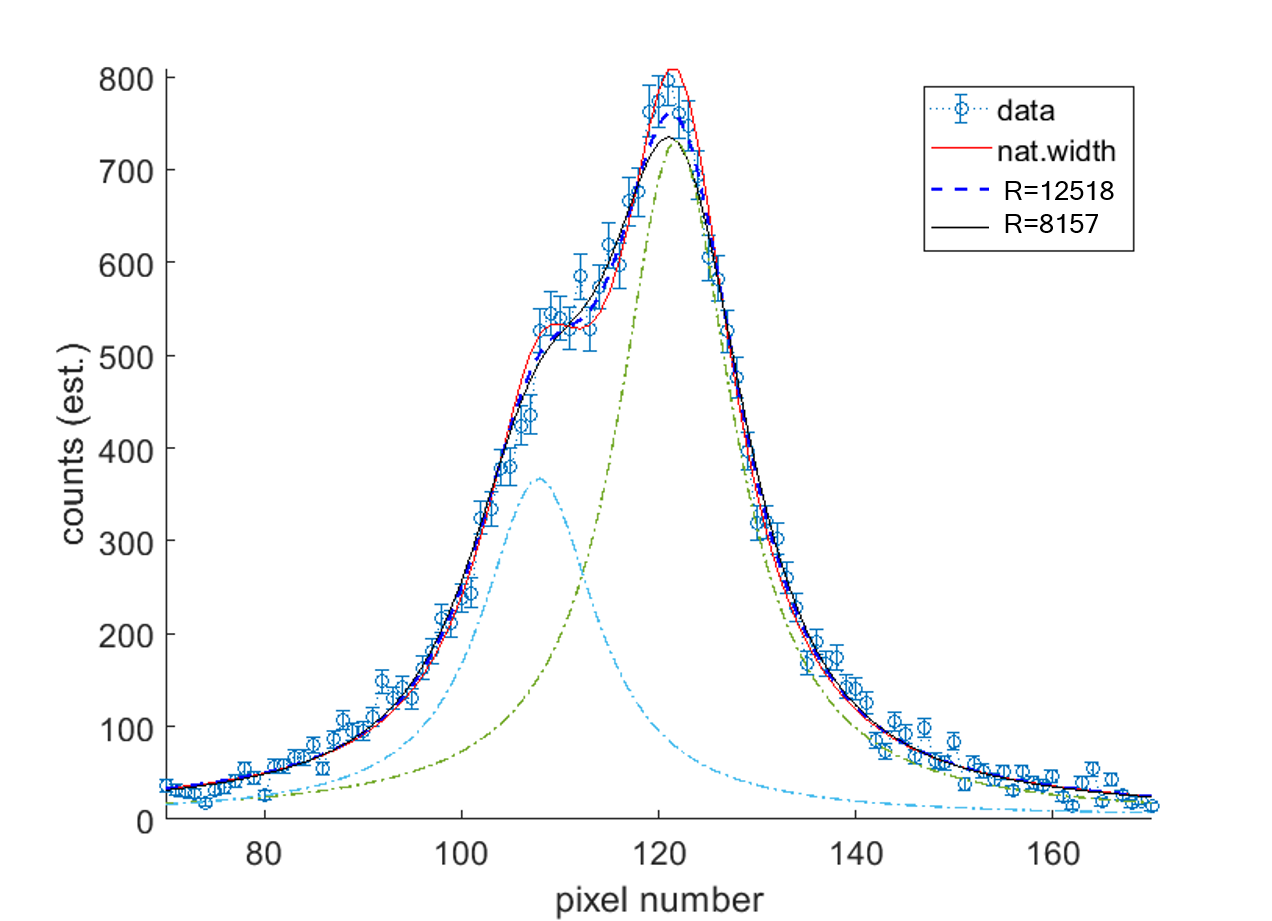}
  \caption{Left: Schematic of CAT grating diffraction.  The grating is inclined by angle $\alpha$ relative to incident \xrays.  For soft \xrays reflection from the grating bar sidewalls enhances DE for orders where $\beta_m \approx -\alpha$.  Towards harder \xrays ($E \gtrapprox 1$ keV for Si) the grating becomes increasingly transparent, and a microcalorimeter could provide high energy resolution at the telescope focus.  Right: Measurement of the Al K$_{\alpha}$ doublet diffracted into 18$^{\mathrm th}$ order through a pair of aligned CAT gratings in a laboratory XGS. The red line is the natural width of the doublet, the dashed line is the best fit to the data, showing that the gratings perform at the $R \approx 10^4$ level, and the black solid line is the curve for the lower $3\sigma$ confidence limit, corresponding to $R = 8157$. The dashed–dotted lines show the individual K$_{\alpha 1}$ and K$_{\alpha 2}$ components with their natural widths on top of the weak sloped background. From Ref.~\cite{ApJ}. }
  \label{fig:CAT}
\end{figure}

The \emph{critical-angle transmission (CAT)} grating was designed to meet these desires across a broad soft \xray band \cite{OE,AO,ApJ,dphWSPC}.  It acts like a blazed TG, reflecting incident \xrays at grazing angles from the sidewalls of freestanding, ultra-high aspect ratio grating bars (see Fig.~\ref{fig:CAT}).  Despite using reflection to generate a blazing effect, CAT gratings are true TGs as defined at the beginning of  Sect.~\ref{sec:TG}, and follow Eq.~\ref{eq:ipge}.

The reflection produces wavefronts at the grating exit plane that are tilted in the blaze direction 
($f(x) = \exp{\left( i\phi (k)x/p \right)}$ across gap width $a$; see also Fig.~\ref{fig:groovefunction}), similar to the case of reflection gratings with sawtooth profiles (see \cref{sec:RG}).  Effectively, the grating groove diffraction envelope is shifted to the blaze direction, enhancing efficiency in higher diffraction orders.  
Most incident \xrays travel through vacuum without absorption.  The basic DE characteristics of CAT gratings can be explained with the approximations from \cref{sec:diffgeom}, using a groove function with a constant phase slope for the gaps between grating bars, and an amplitude based on sidewall reflectivity \cite{OE,AO}.

The grating bar sidewalls (GBSW) are inclined relative to the incident \xrays by an angle below the critical angle of total external reflection (TER), $\theta_c(k) \approx \sqrt{2 \delta_n (k)}$ (where $1 - \delta_n (k)$ is the real part of index of refraction, $n$) of the grating bar material \cite{Attwood17}.  Therefore, \xrays reflect efficiently from the GBSWs, and diffraction orders near the angle of specular reflection ($\beta_m \approx -\alpha$) are enhanced, or ``blazed" (see Fig.~\ref{fig:CAT}).  $\theta_c$ is on the order of 1-3$^{\circ}$ for soft \xrays.  (For typical designs, orders $m \approx $ 7-9 are blazed for $\lambda = 1.5$ nm).  Small graze angles lead to a challenging CAT grating design with long and narrow grating bars of aspect ratio $AR = d/b \sim 50-150$ (see Fig.~\ref{fig:TG_SEMs}).

Most CAT gratings to date have been fabricated with $p = 200$ nm and a bar height of $d = 4$ $\mu$m.  The grating bars can be as narrow as 40-50 nm. (For grating bars with such a high aspect ratio one often refers to $d$ as grating bar height instead of grating thickness.)  Similar to the \chan LETGs, CAT gratings lack a supporting (and absorbing) membrane.  Instead, they feature an integrated (L1) cross-support mesh with 5 $\mu$m period, and a (L2) coarse hexagonal mesh with a 1 mm pitch.  All three features are simultaneously created from the device layer (front side) of a silicon-on-insulator (SOI) wafer using deep reactive-ion etching (DRIE).  The $\sim 0.5$ mm thick handle layer is DRIE'd with a hexagonal pattern that matches the front side L2 hexagons.  The GBSWs are aligned parallel to the vertical \{111\} planes of the $<$110$>$ device layer.  Since DRIE leaves rough sidewalls, this crystal orientation is used to ``polish" the GBSWs post-DRIE through immersion in KOH solution \cite{alex2}.  The gratings have to be dried in a critical-point dryer to prevent stiction due to liquid-vapor surface tension.  Atomic layer deposition can be used to coat CAT grating bars with thin films of higher electron density such as Pt, thereby increasing $\theta_c$ or extending the useful bandpass toward shorter wavelengths \cite{SPIE2016}.

CAT gratings are routinely fabricated at a size of $\sim 32\times 32$ mm$^2$ (see Fig.~\ref{fig:TGfam}).  At this size, large spectrographs require hundreds up to about 2000 gratings.  A volume production approach, using 4X optical projection lithography with electron-beam written masks as the initial patterning step \cite{OPL}, has been demonstrated, where tens of gratings could be fabricated at once from a single 200 mm SOI wafer \cite{SPIE2020}.  Larger grating sizes are being explored. At some size they will require bending and/or period chirping, depending on $R$ and $A_\mathrm{eff}$ requirements \cite{GuentherSPIE2020}.

The high $AR$ of CAT grating bars requires going beyond the thin grating approximation and scalar diffraction theory for accurate DE modeling.  Instead, full EM vector-based rigorous coupled-wave analysis (RCWA) \cite{RCWA} can be used.
Measured DE of 4 $\mu$m deep etched CAT gratings is in the range of 80-100\% of RCWA model predictions \cite{AO,ApJ}.  Differences often can be modeled with the addition of a Debye-Waller-type roughness factor \cite{ApJ,SPIE2020}.  Deeper 200 nm-period gratings ($d = 6$ $\mu$m, see Fig.~\ref{fig:TG_SEMs}) are under development with increased DE compared to less deep gratings (DE $> 40$\% summed over blazed orders, see Fig.~\ref{fig:TG_DE}) \cite{SPIE2023}.  The camera of a CATXGS collects multiple diffraction orders from a range of wavelengths.  At many wavelengths, multiple orders are blazed and contribute to $A_\mathrm{eff}$ and an effective resolving power can be defined as the efficiency-weighted sum of $R$ over the contributing orders \cite{Guenther2017,LynxXGS}.  
The detector must have enough intrinsic energy resolution to separate spatially overlapping orders with differing $\lambda$ \cite{GuentherSPIE2019}.

Due to their ability to blaze to high diffraction orders, CAT grating spectrographs can provide very high $R$ in the soft \xray band.  Experiments using sub-apertured focusing optics with $\sim 1$ arcsec line-spread function (LSF) and individual \cite{SPIE2016,AO2019}, as well as pairs of CAT gratings \cite{ApJ} have demonstrated $R > 10^4$ at Al-K$_{\alpha}$ wavelengths ($\lambda = 0.834$ nm, see Fig.~\ref{fig:CAT}).  These results also show that CAT gratings are fabricated with very small period variations, $\Delta p$, within and between gratings ($\Delta p/p < 1/R$).  Examining an example FOM for CAT grating technology in Fig.~\ref{fig:TG_DE} shows gains on the order of a conservative factor of five compared to \chandra gratings.

\section{Reflection Gratings}\label{sec:RG}

Blazed RGs with sawtooth surface reliefs are beneficial for \xray astronomy due to their ability to concentrate DE in a specified bandpass through choice of groove-facet angle, thereby maximizing throughput for sensitive spectroscopy \cite{Loewen97,McEntaffer_DeRoo21}. 
Simultaneously, aberration-correcting, chirped groove layouts can be implemented to achieve high $R$ in a Wolter-type XGS. 
Many grating facets can be built into modular arrays to attain high $A_\mathrm{eff}$, and hence a large FOM for weak absorption line detection. 
Compared to TGs, this technology offers higher DE and provides greater flexibility in groove profile modification for blazing.

Grating geometry is constrained by the physical requirement that the graze angle on the groove facets, $\zeta$, must be smaller than the relevant critical angle for TER, $\theta_c (k) \approx \sqrt{2 \delta_n (k)}$, which is typically a few degrees in the soft \xray bandpass. 
In this regime, dense materials with high atomic number lead to increased reflectivity over an extended bandpass with larger $\theta_c$; only thin layers are required owing to the $1/\mathrm{e}$ attenuation depth being on the order of nanometers \cite{Attwood17}.
Additionally, low facet roughness, $\sigma$, is desired to limit scatter and absorption \cite{Gibaud2009}, with the Fraunhofer criterion for a smooth surface, $\lambda > 32 \sin \left( \zeta \right) \sigma$, 
indicating nanoscale $\sigma$ \cite{Beckmann63}. 
As a consequence of their use at grazing incidence, many grating facets, each with a substantial size, must be stacked and aligned into modules to achieve sufficient $A_\mathrm{eff}$ for sensitive spectroscopy. 
Manufacture strategies typically center on replica production from a master template due to the difficulties and costs associated with fabricating custom blazed gratings \cite{RGS,LynxOPG}. 

Blazed spectral response from an RG with a groove-facet angle $\delta$ can be described approximately in terms of scalar DE for an infinite sawtooth phase grating. 
Using the framework from \cref{sec:diffgeom}, the groove function defined by \cref{eq:groove_func} can be written as $f(x) = \exp{\left[i \phi (k) x / p \right]}$ for $0 < x \leq p$, where $\phi (k)$ is the phase shift for an ideal groove facet of depth $d = p \tan \left( \delta \right)$. 
DE for the $m^{\text{th}}$ order comes out to
\begin{equation}
    \frac{I^{(m)}}{I_0} = \sinc^2 \left( \frac{\phi (k)}{2} - m \pi \right) ,
\end{equation}
where $\phi (k) = k d \sin \left( \gamma \right) \left[ \cos \left( \alpha \right) + \cos \left( \beta_m \right) \right]$. 
It can be shown that DE for $m>0$ is maximized when $\beta_m$ matches $\beta_b \equiv 2 \delta - \alpha$, similar to reflection from a mirror tilted at an angle $\delta$ with an incidence angle $\alpha$.
\begin{figure}
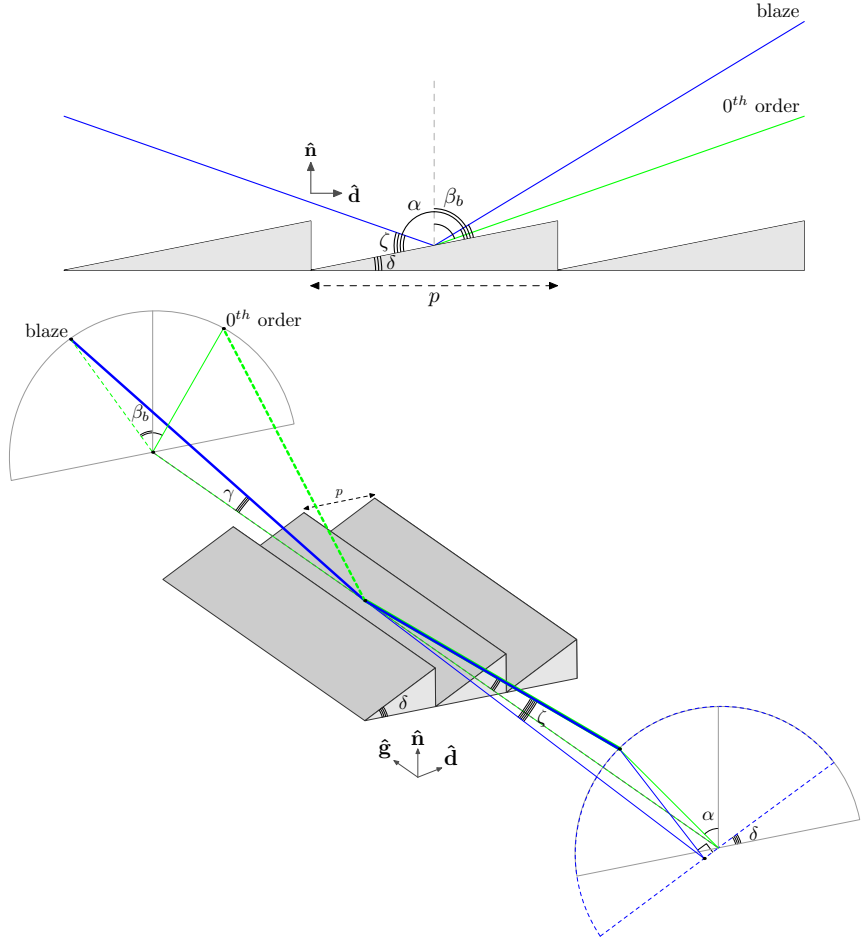

\centering
\includegraphics[scale=0.82]{figures/diffraction_arc54.mps}
\includegraphics[scale=0.47]{figures/diffraction_arc6.mps} 
\caption{ A geometric ray incident on a blazed RG. With incidence angles $\alpha$ and $\gamma$ as in \cref{fig:general_conical}, the ray strikes an ideal sawtooth facet with $\delta$ at an angle $\zeta$ relative to its surface such that $\sin \left( \zeta \right) = \sin \left( \gamma \right) \cos \left( \delta - \alpha \right)$. Illustrated for the special case of in-plane diffraction with $\gamma = 90^{\circ}$ (top) and an arbitrary case of conical diffraction with generic $\gamma$ (bottom), the grating blaze in either case is associated with $\beta_b = 2 \delta - \alpha$ such that scalar DE is maximized when $\beta_m = \beta_b$.}\label{fig:blaze_RG} 
\end{figure}
Inserting $\beta_m = \beta_b$ in \cref{eq:general_GE} yields $m \lambda_b = 2 p \sin \left( \zeta \right) \sin \left( \delta \right)$ for $m>0$, where $\sin \left( \zeta \right) = \sin \left( \gamma \right) \cos \left( \delta - \alpha \right)$. 
Illustrated geometrically in \cref{fig:blaze_RG}, this describes a discrete set of blaze wavelengths, $\lambda_b$ with $m = 1, 2, 3 \dots$, for which DE is expected to be maximized according to scalar diffraction. 
More accurate modelling, however, usually requires full EM vector-based treatments such as RCWA \cite{RCWA} or the integral method \cite{Goray10}. 

Chirped groove layouts can be characterized by a grating vector, $\mathbf{K} (x,z)$, that varies spatially over the surface of a flat RG with $\mathbf{\hat{n}} = \mathbf{\hat{y}}$.
The types of gratings that can be realized with efficient, blazed groove facets patterned over groove layouts prescribed for aberration correction depend on the fabrication technology pursued. 
This section describes how \xray RG fabrication methods 
have evolved since 
the gratings for the \xmm/\rgs were manufactured, with current efforts employing nanofabrication techniques based on electron-beam (e-beam) lithography 
for master grating fabrication and nanoimprint lithography 
for replica production. 
Additionally, alignment approaches for modular arrays have evolved and are summarized as well. 

\subsection{The \xmm Reflection Gratings}
\label{subsec:RGS}

The \rgs of \xmm is the first space-borne instrument to implement RGs in a Wolter-type XGS 
\cite{Mao23}. 
Two identical \rgs instruments are incorporated into two of the three Wolter-I telescopes on the \xmm payload. 
Each set of focusing optics consists of 58 nested mirror pairs with shell diameter ranging from about 0.3 to 0.7~m to produce an overall PSF of 5-6 arcsec FWHM and 14-15 arcsec half-energy-width over a 7.5~m focal length. 
Both \rgs instruments house a Reflection Grating Array (RGA) placed near the exit of the telescope optics, $\sim$6.7~m from the telescope focus, which nominally features 182 identical gratings aligned into modules to disperse soft \xrays with $\lambda \approx 5-35$~\AA~ for spectroscopy. 
Dispersed spectra are imaged by a focal plane camera with nine charge-coupled devices (CCDs), where order separation is carried out using the intrinsic energy resolution of these detectors.  
This is enabled by a geometry where the center of each grating, along with the telescope focus and the diffracted focus, lies on the Rowland torus \cite{RGS,Brinkman96,Kahn96,Paerels10}.

Inside each RGA are replicas of a master grating fabricated via mechanical ruling engine, with each produced by transferring the ruled pattern into a $\sim$25~$\mu$m-thick layer of epoxy on a 1~mm-thick SiC substrate before being coated with $\sim$200~nm of Au for reflectivity \cite{Rasmussen98,RGS}. 
These gratings were designed for a \emph{grazing in-plane geometry (GIPG)} defined by $\sin \left( \gamma \right) = 1$ in \cref{eq:general_GE} with $\alpha \approx 88.4^{\circ}$ and $\mathbf{\hat{d}}$ pointing toward the focal plane. 
A very shallow groove-facet angle of $\delta \approx 0.7^{\circ}$ allows for the groove-facet graze angle, $\zeta = 90^{\circ} - \alpha + \delta \approx 2.3^{\circ}$, to be smaller than $\theta_c$ of Au. 
This results in $m \lambda_b \approx 2 p \zeta \delta$ for $m>0$ such that 
$p \approx 10^3 m \lambda$ is required for blazed diffraction at an angle $\beta_b = 2 \delta - \alpha \approx -87^{\circ}$. 
Each grating is approximately 200~mm long in the direction of $\mathbf{\hat{d}}$ and 100~mm wide along $\mathbf{\hat{g}}$. 
Strengthening ribs made of Be aid to maintain flatness to one optical wavelength (634.8~nm) along $\mathbf{\hat{d}}$ \cite{Brinkman96}. 

The groove density, $1/p$, on each grating replica varies along $\mathbf{\hat{d}}$ such that the direction of $\mathbf{K}$ stays fixed while its magnitude changes to account for $\alpha$ changing with position on the grating from the converging telescope beam \cite{RGS}. 
Defining $x_{\text{min}} \leq x \leq x_{\text{max}}$ as the length of the grating along $\mathbf{\hat{d}} \equiv \mathbf{\hat{x}}$ with $x_{\text{min}} \equiv -100~\text{mm}$ and $x_{\text{max}} \equiv 100~\text{mm}$, $|\mathbf{K}| \equiv K(x)$ can be determined using $K(x_{\text{max}})/K(x_{\text{min}}) \approx \left[ \pi/2 - \alpha (x_{\text{max}}) \right]^2 /\left[ \pi/2 - \alpha (x_{\text{min}}) \right]^2$, which is valid in the limit of grazing incidence \cite{Hettrick83}. 
An approximate functional dependence is quoted 
\cite{denherder94} as
\begin{equation}\label{eq:XMM_VLS}
    K(x) \approx \frac{2 \pi}{p_{\text{cen}}} \left[ 1 + \frac{x}{R_c}  \right] \quad \text{for } -100~\text{mm} \leq x \leq 100~\text{mm} ,
\end{equation}
where $p_{\text{cen}} \approx 1.55$~$\mu$m is the central groove period, consistent with $\lambda_b = 15$~\AA~ for $m=1$, and $R_c = 3352$~mm is radius of the Rowland circle for \xmm \cite{Mao23}.
The gratings were aligned into fanned arrays such that $\alpha$ at the center of each is the same along the Rowland circle using a passive alignment scheme that involved abutting each grating substrate against stainless steel alignment rails and fixing into place using spring clips \cite{Kahn96}.
Constrained by blazed orders with $90^{\circ} + \beta_b \approx 3^{\circ}$, the separation between each grating in an array is $\sim$12.5~mm such that the fraction of \xrays intercepted by the \rgs is approximately $\cos \left( \alpha \right) / \cos \left( \beta_b \right) \approx 0.53$ \cite{Brinkman96}. 
Performance metrics for \xmm are summarized in \cref{subsubsec:perf}. 


\subsection{Current Status of Reflection Gratings}\label{subsec:OPG}

Much of the recent and current instrument development for \xray RGs considers an \emph{extreme off-plane geometry (EOPG)} to make advancements in $R$ and spectral sensitivity \cite{McEntaffer_DeRoo21}, though GIPGs similar to the \xmm/\rgs are pursued as well \cite{Savage_2023}. 
An EOPG is characterized by a small $\gamma$, which allows for high DE in high $|m|$ with appropriate choice of $\delta$, which in turn leads to high $R$ in a bandpass of interest \cite{Cash91}. 
With $\zeta$ also small by $\sin \left( \zeta \right) = \sin \left( \gamma \right) \cos \left( \delta - \alpha \right)$ (see \cref{fig:blaze_RG}), $\alpha$ can match $\delta$ in a \emph{Littrow configuration} that maximizes DE with $\alpha = \beta_b = \delta$ and $\zeta = \gamma$. 
The blaze wavelength for this scenario is $m \lambda_b \approx 2 p \gamma \sin \left( \delta \right)$ for $m>0$, showing that $p \gtrapprox 10 \csc (\delta) m \lambda$ is required for blazed diffraction at a typical graze angle $\zeta < \theta_c$. 
This indicates $p \gg \lambda$ as for the \xmm gratings, but by a smaller margin such that $p$ is typically a few hundred nanometers.
Additionally, this geometry with small $\gamma$ allows for gratings to be packed more closely into modular arrays without obstructing any propagating orders \cite{Cash91}. 

A chirped groove layout for an EOPG that takes into account the converging quality of the incident beam, in contrast to the GIPG discussed in \cref{subsec:RGS}, manifests as having $\mathbf{K}$ that varies in both direction and magnitude. 
An approximate solution is the \emph{radial groove profile}, where $\mathbf{\hat{g}}$ points toward a single point on the focal plane while $\mathbf{\hat{d}}$ points in the azimuthal direction in the grating substrate plane with $\mathbf{\hat{n}} = \mathbf{\hat{y}}$ \cite{Cash83,RasmussenSPIE2004}.  
This scenario of a fanned-groove layout can be expressed as 
\begin{equation}\label{eq:VLS_radial}
    \mathbf{K} (x,z) = 2 \pi \frac{z \mathbf{\hat{x}} - x \mathbf{\hat{z}}}{\theta_p \left( x^2 + z^2 \right)} ,
\end{equation}
where $\sqrt{x^2 + z^2}$ is the radial distance to the groove hub and $\theta_p$ is a small angle that parameterizes the constant angular period for the grating.
This angle $\theta_p$ can be related to the central groove period, 
$p_{\text{cen}} \equiv \theta_p z_{\text{cen}}$, from $\mathbf{K} (0,z_{\text{cen}}) = 2 \pi \mathbf{\hat{x}} / \left( \theta_p z_{\text{cen}} \right)$, where $z_{\text{cen}}$ is the distance to the groove hub at $x=0$. 
A radial groove profile creates a gradient in $p$ across the grating similar to the in-plane case for \xmm, but additionally, the tilted grooves in principle ensure that $\alpha$ is approximately constant across each grating in a converging beam of radiation with dispersion at the focal plane given approximately by $m\lambda /\theta_p$ \cite{Cash83}. 
This holds under the conditions of grazing incidence and a slowly-converging radial profile such that $|x| \ll z$ across the grating, or equivalently, that $z_{\text{cen}}$ is much larger than the grating width centered at $x=0$. 

Following the fabrication of a blazed RG master with radial grooves and subsequent mass replication (discussed in \cref{subsubsec:fab}), a current alignment strategy for replicated gratings is inspired by silicon pore optic (SPO) technology, where ribbed silicon substrates with a precise wedge angle are stacked into modules using direct bonding of silicon surfaces \cite{Collon21}. 
This semi-passive alignment scheme 
makes use of a nanoimprint variant that results in groove facets imprinted in a silica-like resist as described below, which enables grating replicas to be stacked and bonded similar to SPO technology \cite{Donovan21}.
With the success of this approach relying on the fabrication of a master grating with efficient, blazed grooves patterned over a chirped layout, \xray RG fabrication efforts are outlined in \cref{subsubsec:fab} before highlighted performance results from grating prototypes are summarized in \cref{subsubsec:perf}.

\subsubsection{Fabrication}\label{subsubsec:fab}
With heritage dating back to the early $19^{\text{th}}$ century, 
the mechanical ruling process gives rise to a sawtooth-like topography, where the effective $\delta$ depends directly on the shape of the ruling tip \cite{Loewen97}. 
Owing to the development of mechanical ruling engines with interferometric feedback control, 
nanoscale groove placement precision is possible, but the technique comes with the potential for surface roughness impacting DE. 
\xray gratings with radial grooves defined by \cref{eq:VLS_radial} were first proposed to be fabricated via mechanical ruling in the 1980s \cite{Cash83}. 
However, this has not been demonstrated, and additionally, \xray DE measurements of mechanically-ruled gratings report losses in EOPGs \cite{Werner77}.

One of the other fabrication techniques most pursued for \xray RGs 
is interference lithography. 
Also known as holographic recording, such a lithographic process classically uses two beams split from a laser of wavelength $\lambda_0$ to create a sinusoidal interference pattern with a period $p_0 \equiv \lambda_0 / \left( \sin \left( \theta_1 \right) - \sin \left( \theta_2 \right) \right)$, where $\theta_1$ and $\theta_2$ are coplanar angles for each beam \cite{Noda1974,Loewen97}. 
Given sufficiently low photoresist contrast, the recording process postdevelopment results in a topography that is also approximately sinusoidal with $K = 2 \pi / p_0$.

As a result, an RG fabricated in this way is expected to function like a sinusoidal phase grating with $f(x) = \exp{\left(i \phi (k) \left[1 + \sin (K x) \right] / 2 \right)}$ in an ideal case such that DE is distributed among low $|m|$ as described by $I^{(m)}/I_0 = \left|\left|J_m \left( \phi (k) / 2 \right) \right|\right|^2$, where $J_m$ is the $m^{\text{th}}$-order \emph{Bessel function of the first kind} and $\phi (k) = k d \sin \left( \gamma \right) \left[ \cos \left( \alpha \right) + \cos \left( \beta_m \right) \right]$ with $d$ as the sinuosoid depth. 
An approximate blazed grating can be realized by ablating one side of the sinusoidal grooves through directional ion-beam etching \cite{Loewen97}, 
but gratings fabricated in this way have been shown to perform with lower overall DE in the soft \xray \ band, likely due to etch-induced roughness that causes absorption and scatter \cite{Tutt16}. 

Holography can also be used to define a groove layout for crystallographic etching in Si to produce smooth, triangular groove facets defined by $\{111\}$ planes of the crystal when $\mathbf{\hat{g}}$ is aligned with  the $\langle 011 \rangle$ direction on the wafer surface \cite{Franke97,Kahn99}. 
Similar to the first step of the fabrication processes for HETGS on \chan as outlined in \cref{subsec:HEGMEG}, interference lithography is first used to produce a groove layout in high-contrast photoresist that resembles a lamellar profile of lines and spaces. 
Following a series of etch steps, the effective $\delta$ depends on the surface-normal orientation of the wafer such that, for example, a $\langle 100 \rangle$ surface leads to a symmetric sawtooth with $\delta \approx 55^{\circ}$ and $\langle 311 \rangle$ leads to an asymmetric sawtooth with $\delta \approx 30^{\circ}$. 
The topography of such a grating can be inverted through replication via nanoimprint to achieve a sharp groove apex with the flat tops produced by the etch mask laying at the groove base to be shadowed by incident radiation \cite{Chang03}.
Such wet-etching processes are beneficial for producing blazed grooves with exceptionally smooth facets of specified $\delta$, which enable high overall DE and an effective blaze response \cite{Miles18}. 
While gratings with $p \lessapprox 100$~nm can be fabricated using holography in some cases \cite{Chang08}, $\lambda_0$ places a limit on $p_0$. 
Additionally, curved grooves generally are produced when the recording beams are not coplanar \cite{Noda1974} and hence a true radial groove layout defined by \cref{eq:VLS_radial} cannot be produced holographically. 
This process also demands precise alignment between $\mathbf{\hat{g}}$ and $\langle 011 \rangle$ to achieve smooth and continuous groove facets with limited etch defects. 

E-beam lithography has been pursued for \xray RG technology in recent years due to its ability to pattern custom groove layouts with sub-nm precision \cite{Chen15,DeRoo20}. 
The e-beam recording mechanism is based on local manipulation of the average molecular weight in a resist film, which is induced by electron exposure quantified by a lithographic dose \cite{Chen15}. 
With appropriate choice of dose in a standard e-beam process for grating manufacture, molecular-weight-dependent etching of the resist during wet development leads to a lamellar topography over a specified layout. 
This serves as a starting point for multiple possible fabrication approaches for realizing a blazed, sawtooth topography in the underlying substrate. 

\begin{figure}
 \centering
 \includegraphics[scale=0.65]{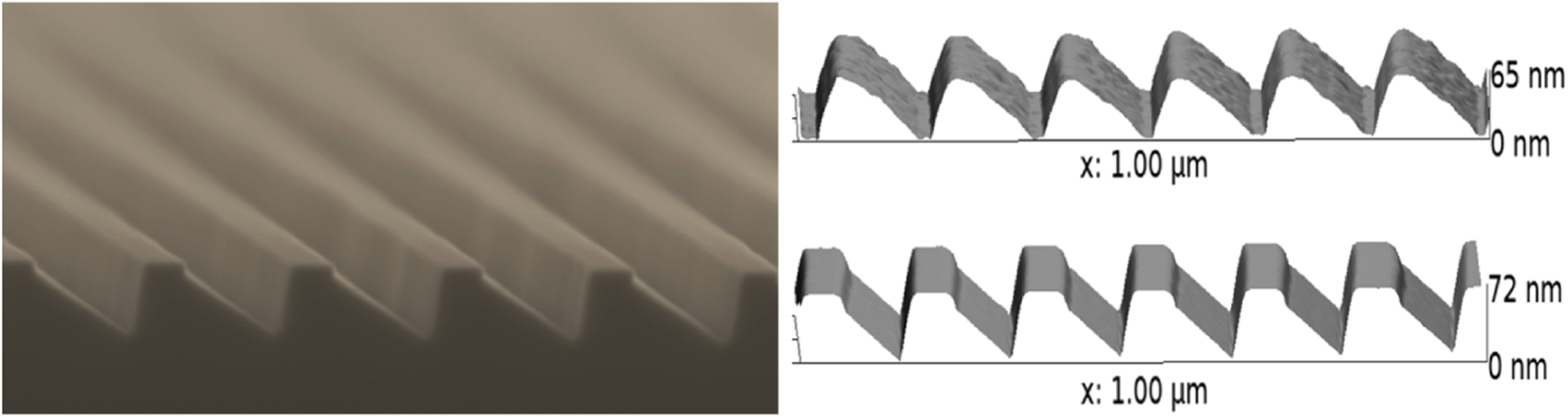}
 \caption{Left: FESEM cross-section image of a $p \approx 160$~nm grating KOH etched in $\langle 311 \rangle$ Si. Right: AFM images of a coated replica (top) and the KOH-etched grating used as a stamp for UV-nanoimprint (bottom) \cite{Miles18}.}\label{fig:KOH_grooves} 
 \end{figure}

One approach is e-beam-defined KOH etching, where interference lithography is replaced in the process described above for groove patterning \cite{DeRoo16,Miles18,Champey22}. 
An example of this is shown in \cref{fig:KOH_grooves} under field-emission scanning-electron microscopy (FESEM) and atomic force microscopy (AFM), where a KOH-etched $\langle 311 \rangle$ Si wafer was patterned with an approximate radial groove layout with $p \approx 160$~nm. 
This resulted in the topography with $\delta \approx 30^{\circ}$ shown  in the bottom AFM image and under FESEM; the Si grating was then used as a direct stamp in a UV-nanoimprint process that inverted the topography, which is shown following a sputter-coat of 5~nm of Cr and 15~nm of Au in the top AFM image \cite{Miles18}. 
Although this technique enables high DE through sharply-defined groove facets, a radial layout with fanned grooves still results in the formation of groove facets that are confined to the cubic crystal structure of Si; this can lead to degradation of $R$ in an XGS. 


An alternative method of grating manufacture that does not rely on substrate etching is \emph{thermally-activated selective topography equilibration (TASTE)}, which combines grayscale e-beam lithography and polymer thermal reflow to produce smooth and continuous surface reliefs in a thermoplastic resist \cite{Schleunitz14}. 
The former is based on dose-modulated e-beam exposure such that multi-leveled features are produced in resist via molecular-weight-dependent etch rates during a timed wet development. 
The latter is based on the dependence of the local glass-transition temperature of the resist on molecular weight following e-beam exposure; such material contrast enables selective equilibration so that, with an appropriate temperature for heat treatment, exposed resist becomes molten while unexposed resist remains in its glass state. 
The final product expected from an ideal TASTE process 
is a sawtooth-like topography that serves as a grating mold with $\delta$ that depends largely on the geometry of the initial grayscale pattern. 
With no dependence on the crystal structure of the substrate, TASTE has the potential for realizing RGs that feature blazed groove facets patterned over a non-parallel, chirped groove layout, thereby enabling high sensitivity and high $R$ in an XGS \cite{McCoy18,McCoy20}. 
\begin{figure}
 \centering
 \includegraphics[scale=0.25]{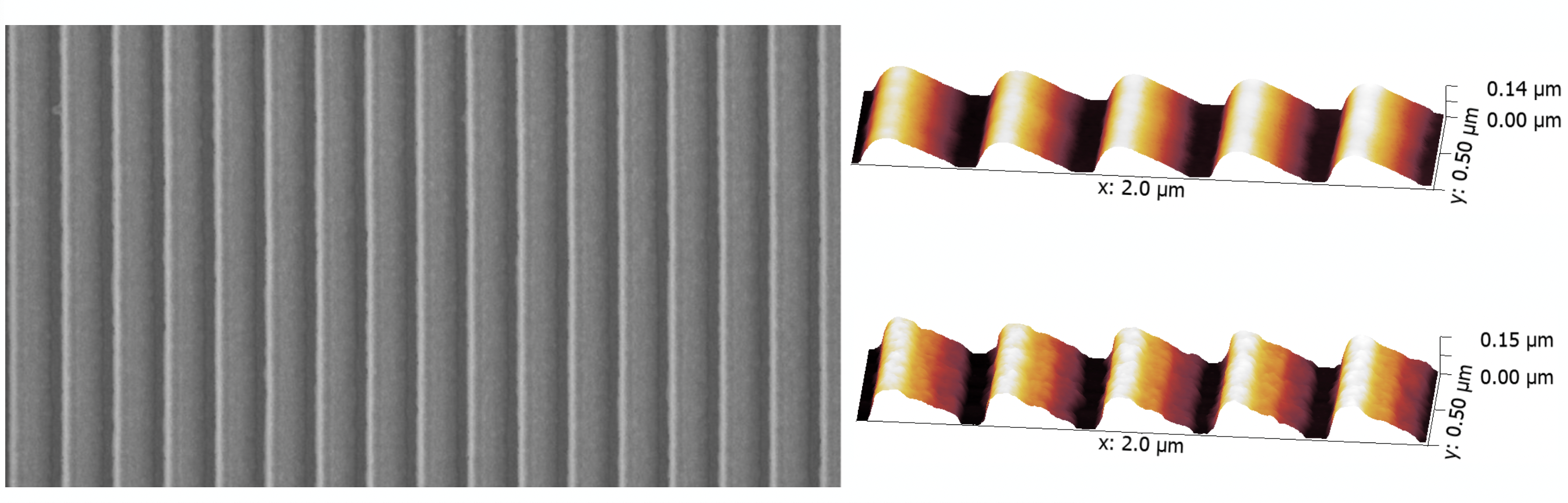}
 \caption{Left: FESEM top-down image of a coated RG with $p = 400$~nm fabricated using TASTE. Right: AFM images of the coated grating grooves with $\delta \approx 27^{\circ}$ (top) and the initial grayscale pattern in e-beam resist (bottom) \cite{McCoy20}.}\label{fig:TASTE_grooves} 
 \end{figure} 
A grating prototype with $p=400$~nm and $\delta \approx 27^{\circ}$ fabricated in $\sim$130-nm-thick poly(methyl methacrylate) e-beam resist via TASTE is shown under FESEM and AFM in \cref{fig:TASTE_grooves}. 

Accompanying these techniques in e-beam lithography is a nanoimprint variant known as \emph{substrate conformal imprint lithography (SCIL)}, which centers on the use of a low-cost, flexible stamp molded from a rigid master template \cite{Verschuuren17}. 
With stamp features carried in a modified form of polydimethylsiloxane that has an increased Young’s modulus relative to that of the standard form, SCIL offers a way for nanoscale patterns to be imprinted in resist over large areas using a stamp that conforms to the substrate surface. 
SCIL is compatible with a brand of inorganic resist that cures through a thermodynamically-driven, silica sol-gel process to realize grating grooves imprinted in a silica-like material \cite{Verschuuren17,McCoy20b}. 
Automated production platforms are capable of producing $\gtrapprox 700$ imprints in this resist at a rate of $\sim$60, 150-mm-diameter wafers per hour, without pattern degradation \cite{Verschuuren17}. 
Overall, SCIL enables the SPO-inspired modular grating arrays described in \cref{subsec:OPG} to be manufactured, with the silica sol-gel resist providing a bonding surface for ribbed Si substrates \cite{Donovan21}. 



\subsubsection{Performance}\label{subsubsec:perf}

The \rgs of \xmm, described in \cref{subsec:RGS}, provides 
$R = \lambda / \Delta \lambda \approx 150 - 800$ over $\lambda \approx 5-35$~\AA~ with $\Delta \lambda$ taking into account the blurred telescope PSF, alignment and flatness errors, aberrations in the chirped groove layout defined by \cref{eq:XMM_VLS}, detector spatial resolution, as well as defocus errors \cite{Brinkman89}. 
Effective area, $A_\mathrm{eff}$, of the instrument peaks at $\sim$150~cm$^2$ for $\lambda_b$ at $m=1$ with the DE of individual gratings measuring, on average, 10\%-18\% for $m=1$ and 2\%-6\% for $m=2$, with scatter contributing to about 10\%-15\% of losses \cite{Rasmussen98}. 
Owing to the technological developments in RG fabrication outlined in \cref{subsubsec:fab}, significant improvements in the performance metrics of $R$ and DE have been made in recent years. 
This section highlights test results for e-beam-written RGs that demonstrate these advancements. 

A lamellar RG was patterned via e-beam lithography to a radial groove layout defined by \cref{eq:VLS_radial} for the purpose of testing for $R$ in a Wolter-I telescope system at the \textsc{PANTER} \xray Test Facility, where a $\sim$120~m-long beamline provides quasi-collimated \xrays generated by an electron-impact source and test chamber suited for manipulating optical components \cite{Donovan20}. 
The grating design was parameterized by having a central period of $p_{\text{cen}} = 157.57$~nm that converges over a distance $z_{\text{cen}} = 3250$~mm such that $\theta_p \approx 0.01$~arcsec. 
This e-beam pattern extended 100~mm along $\mathbf{\hat{z}}$ and 86~mm along $\mathbf{\hat{x}}$ to be large enough to encompass a converging telescope beam 
before it was dry etched into a Si substrate and then sputter coated with 15~nm of Au for reflectivity at Mg-K$\alpha$ and 5~nm of Cr for adhesion of Au to Si. 
A single pair of mono-crystalline Si, Wolter-I optics \cite{Zhang18} with a $\sim$8.4-m focal length and a 156-mm radial intersection node was used for testing, with each segment being $\sim$100 mm long and spanning $30^{\circ}$ of a full shell \cite{Donovan20}. 
These optics served to focus slightly-diverging, Mg-K$\alpha$ \xrays over a distance of $\sim$9.1 m, where a CCD 
was used to sample the optic PSF, as well as the diffracted arc produced by the lamellar grating placed to intercept the converging beam at a position consistent with $z_{\text{cen}}$. 

Following a series of alignment steps to establish an EOPG with $\alpha \approx 30^{\circ}$ and $\gamma \approx 1.73^{\circ}$ \cite{Donovan20}, LSFs for $m=0$ and $m=5$ were imaged and analyzed using 
$\lambda \approx 9.89$~\AA~ associated with the spectral-line doublet from Mg-K$\alpha$ fluorescence. 
\begin{figure}
 \centering
  \includegraphics[scale=0.17]{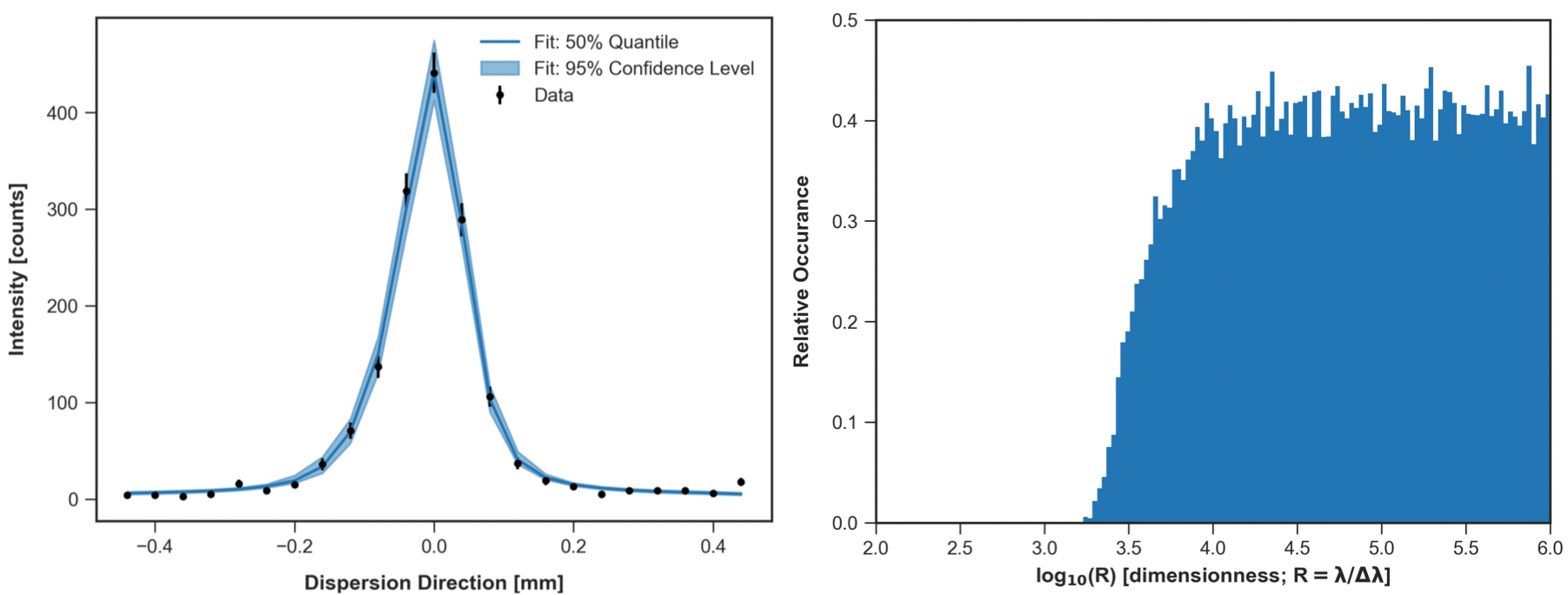}
 \caption{Analysis of an RG tested at PANTER for $R = \lambda / \Delta \lambda$.  Left: Measured $m=5$ LSF fitted using Bayesian modelling with $\Delta x \approx 106$~$\mu$m. Right: Posterior probability distribution for $R$ associated with the grating alone indicating $R>4500$ \cite{Donovan20}. }\label{fig:panter1} 
 \end{figure} 
A directly-measured LSF width for $m=5$, with fitted data shown in \cref{fig:panter1}, led to $\Delta x \approx 109 \mu$m with $R = x / \Delta x \approx 936$ for the overall system, where $x = r \left[ \sin \left( \alpha \right) + \sin \left( \beta_5 \right) \right] \approx 102$~mm with $\beta_5 \approx 33^{\circ}$ is the dispersion distance with $r = L \sin \left( \gamma \right) = 98.3$~mm as the diffracted-arc radius and $L=3251.49$~mm as the system throw \cite{Donovan20}. 
However, Bayesian modeling of the $m=5$ LSF via Markov chain Monte Carlo sampling and analysis of a resulting posterior probability distribution, shown in \cref{fig:panter1}, demonstrated with 94\% confidence that $R$ associated with the grating alone is $> 4500$. 
This LSF model takes into account contributions from the optic PSF, source size, natural line broadening and substrate curvature such that $R > 4500$ is characteristic of groove-period and diffraction errors associated with the grating \cite{Donovan20}. 
Indicating that high $R$ can be achieved in an improved Wolter-I system with a tighter tolerance for substrate flatness, these results, coupled with an interferometric analysis of a RG fabricated in similar way \cite{DeRoo20}, demonstrate that e-beam lithography is capable of patterning chirped groove layouts suitable for an XGS. 

TASTE, described in \cref{subsubsec:fab}, provides an avenue for achieving a blazed, sawtooth-like topography using e-beam lithography directly \cite{Schleunitz14,McCoy18}.
This technique was first applied to \xray RG technology for the fabrication and subsequent DE synchrotron testing at the Advanced Light Source (ALS) of Lawrence-Berkeley National Laboratory \cite{McCoy20}. 
\begin{figure}
 \centering
 \includegraphics[scale=0.9]{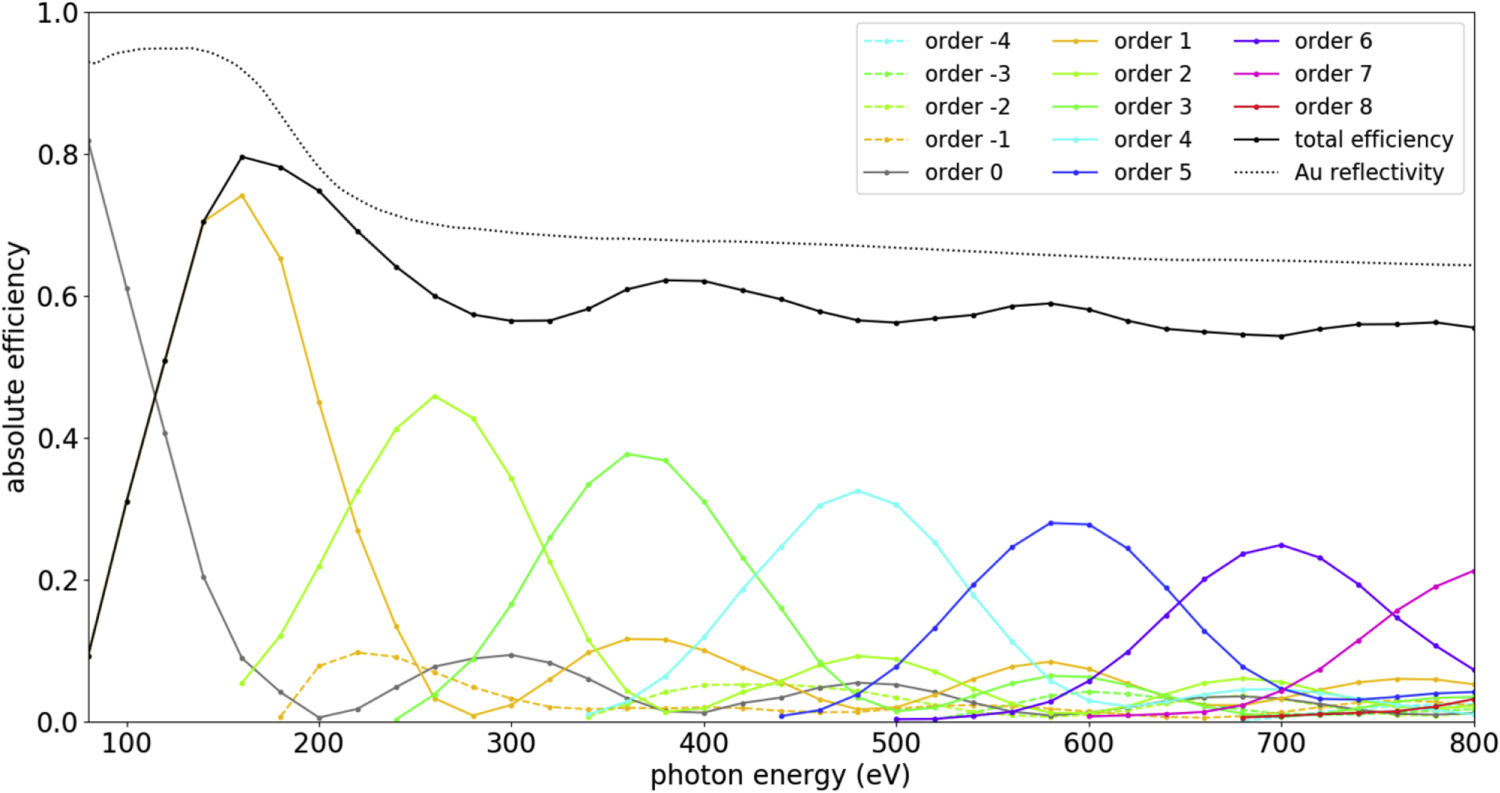}
 \caption{Absolute DE ($m = -4$ to $8$) as a function of 1240~(eV~nm)$/\lambda$, of the Au-coated TASTE RG shown in \cref{fig:TASTE_grooves} with Fresnel reflectivity for Au at $\zeta \approx 1.7^{\circ}$ plotted for comparison \cite{McCoy20}.}\label{fig:TASTE_eff} 
 \end{figure} 
The grating-prototype grooves shown in \cref{fig:TASTE_grooves} with a uniform period of $p=400$~nm were patterned over 50~mm along $\mathbf{\hat{g}}$ and 7.5~mm along $\mathbf{\hat{d}}$ to encompass a grazing-incidence beam of \xrays produced by a monochromator station at the ALS. 
Following a 15-nm-thick coating of Au achieved via e-beam physical vapor deposition using 5~nm of Ti for adhesion to the resist, shown in \cref{fig:TASTE_grooves}, the grating was tested for DE across 15.5~nm $\gtrapprox \lambda \gtrapprox$ 1.55~nm in an EOPG defined by $\alpha \approx 23^{\circ}$ and $\gamma \approx 1.7^{\circ}$. 
The results shown in \cref{fig:TASTE_eff} demonstrate that the prototype behaves approximately as a blazed grating; total losses relative to the Au coat range from 4\%-12\%, which is attributed to absorption and scatter from groove facets with $\sigma \approx 1.5$~nm RMS \cite{McCoy20}. 
Overall, these results show that TASTE is a promising technique for \xray RG manufacturing.  
Current research focuses on improving the fabrication process and testing a radial TASTE grating for $R$ and DE.


\section{Scientific Investigations Requiring New Grating Technology} \label{sec:newsci}  

Higher resolving power in \xrays than is currently possible with \chan
or \xmm ($R < 1000$) would open the door to many new areas of
investigation.  Resolving thermal lines widths requires $R$ of
$2000$--$4000$, which would provide a way to measure the ion
temperature and not just electron temperature.  Detection of weak
unresolved absorption lines is much improved with higher resolving
power because any given small spectral region ($\Delta \lambda$) has a
contribution from an unresolved absorption line plus the continuum.
There are
weak lines in crowded spectral regions which are potentially important
plasma diagnostics such as dielectronic recombination satellite lines
which provide an additional temperature diagnostic.  In general, higher effective area 
is also needed along with the higher resolving power to compensate for having a 
lower density of photons on the detector (see Section~\ref{sec:Miss} and references
therein for detailed trade studies).
A few areas where higher resolving power will increase sensitivity to
plasma diagnostics are:
\begin{itemize}

\item sensitive searches for weak absorption lines in distant objects
  to study the fundamental baryon content of the universe \citep{brenneman:al:2016,kaastra:2017,mathur:al:2021,nicastro:al:2018};

\item detection of absorption lines in the Galactic halo, for composition, and
  dynamics \citep{brenneman:al:2016,das:mathur:al:2021};

\item study of accretion flows in young stars, relevant to stellar
  evolution and growth of planetary systems \cite{brickhouse:al:2010, gunther:al:2007};

\item resolving the detailed shape of stellar wind profiles, to better
  understand mass loss, and implications for feedback of matter and
  energy \cite{owocki:cohen:2001, gunderson:al:2022}, and resolve
  possible \xray absorption components from stellar winds, as done in
  the ultraviolet \cite{massa:al:2015};

\item resolve the dynamics in extreme objects: measure outflows in
  active galactic nuclei to high precision, or accretion flows and
  jets around black holes and neutron stars \citep{danehkar:nowak:al:2018,kaastra:2017}.

\end{itemize}


\section{Approved and Proposed Missions}
\label{sec:Miss}

XGS instruments were last put in orbit in 1999, but \xray diffraction grating technology has made significant strides since.  Numerous new XGS-carrying missions have been proposed to NASA \cite{Con-X,IXO,Smart-X,OPGWhimExIXO,Bautz}.  A next-generation \xray observatory concept named \lynx was proposed to the 2020 Decadal Survey and recommended as a feasible candidate for a future competition \cite{Lynx}.  Both CAT grating and  off-plane reflection-grating instruments were proposed as part of the \lynx instrument suite \cite{LynxXGS,LynxOPG}, each providing $A_\mathrm{eff} > 4000$ cm$^2$ and $R > 5000$ in the soft \xray band, which would surpass currently orbiting instrument FOMs by factors of 25-50.  Both version would feature large retractable grating arrays similar to \chandra.  \lynx would also feature a powerful microcalorimeter instrument and a large wide-field camera, fed by a $> 1$ m$^2$ effective area, 0.5 arcsec PSF mirror array.

\arcus is a smaller CATXGS mission proposed at the midsize Explorer and Probe levels \cite{Arcus,ArcusProbe}.  It uses aggressive sub-aperturing \cite{Cash91,SPIE2010} in four parallel channels to reduce the PSF in the dispersion direction and combines four spectra onto two readout cameras.  $A_\mathrm{eff}$ is about $400$ cm$^2$, with $R \sim 3000$ \cite{GuentherSPIE2023}.

A number of suborbital rocket missions have incorporated reflection-grating XGS instruments, although for diffuse spectroscopy with $R \sim 10-100$ \cite{2008ApJ...680..328M,2011ExA....31...23O,2019JATIS...5d4006M,2019SPIE11118E..0BM,2020ExA....49....1R}. 
An upcoming rocket mission, the \emph{Off-plane Grating Rocket Experiment (OGRE)}, has requirements of $R > 1500$ ($R > 2000$ goal) and $A_\mathrm{eff} > 20$ cm$^2$ (peaks at $A_\mathrm{eff} > 50$ cm$^2$) across the $10-52$~\AA~ bandpass of interest \cite{Donovan21,Donovan20}. OGRE utilizes polished-Si optics, an array of blazed, radial gratings replicated from a TASTE-recorded master, and soft-\xray-sensitive, electron-multiplying CCDs \cite{Zhang18,2022SPIE12191E..15E}. 
Future launches have been scheduled to observe Capella and exploit the emission-line-rich source to demonstrate high-resolving-power grating spectroscopy.


In-plane RGs have also been employed successfully on a sounding rocket for solar x-ray spectroscopy \cite{Savage_2023}.

CAT gratings, in combination with laterally-graded multilayers, are expected to launch on the selected suborbital \emph{Rocket Experiment Demonstration of a Soft X-ray Polarimeter (REDSoX)} in 2027 to measure the polarization of an active galaxy or a pulsar in an X-ray binary in the 0.2-0.4 keV band \cite{REDSoX}.



%
%
%



\bibliographystyle{spbasic}    
\bibliography{dph,jam}             





\end{document}